\documentclass[a4paper,11pt,reqno]{amsart}
\usepackage[body={16cm,24cm}]{geometry}

\usepackage{amssymb}  
\usepackage{amsmath}  
\usepackage{latexsym} 
\usepackage{graphicx}
\usepackage{hyperref}

\newtheorem{theorem}{\bf Theorem}[section]

\newtheorem{prop}{\bf Proposition}[section]

\newtheorem{rem}{\bf Remark}[section]

\renewcommand{\Im}{\operatorname{Im}}
\renewcommand{\Re}{\operatorname{Re}}

\begin{document}


\title[Periodic finite-band solutions by  Fokas method]{Periodic finite-band solutions 
to the focusing nonlinear Schr\"{o}dinger equation by the Fokas method: inverse and direct problems}

\author{
Dmitry Shepelsky$^{1,2}$, Iryna Karpenko $^{1,3}$, Stepan Bogdanov$^{4}$, Jaroslaw E. Prilepsky$^{4}$
\\ \\
Published in:
Proc. R. Soc. A 480, 20230828
(doi.org/10.1098/rspa.2023.0828)
}

\address{$^{1}$ B. Verkin Institute for Low Temperature Physics and Engineering, Kharkiv, Ukraine\\
$^{2}$V. N. Karazin Kharkiv National University, Kharkiv, Ukraine\\
$^{3}$ University of Vienna, Vienna, Austria\\
$^{4}$ Aston Institute of Photonic Technologies, Aston University, Birmingham, UK
}




\maketitle

\begin{abstract}
We consider the Riemann--Hilbert (RH) approach to the construction of 
periodic finite-band solutions to the focusing nonlinear Schr\"odinger (NLS) equation.
An RH problem for the solution of the finite-band problem has been recently derived via the Fokas method \cite{DFL21, FL21}.
Building on this method, a finite-band solution to the NLS equation can be given in terms of the solution of an associated RH problem, the jump conditions
for which are characterized by specifying the endpoints of the arcs defining the contour of the RH problem and the constants (so-called phases) involved in the jump matrices.
In our work, we solve the problem of retrieving the phases given the solution of the NLS equation evaluated at a fixed time. Our findings are corroborated by numerical examples of phases computation, demonstrating the viability of the method proposed.


\end{abstract}



\section{Introduction}

About fifty years ago, the inverse scattering transform (IST) method was introduced. This method allows us to solve certain one-dimensional nonlinear evolution equations, called integrable equations, on the entire line of a spatial variable. This is achieved by analysing the corresponding linear problems constituting the associated Lax pair
\cite{IST01,IST02,ablowitz1974inverse, ablowitz1991solitons}. 

The IST method can be divided into three main steps.
(i) First, given the initial data, it involves solving a linear auxiliary set of equations and establishing a spectral (direct) problem. This spectral problem maps the initial data to a set of special quantities known as spectral data.
(ii)
Next, the method focuses on understanding the evolution of these spectral characteristics over time.
(iii)
Finally, it tackles an inverse problem, which allows retrieving the partial differential equation (PDE) solution at any desired value of time (evolution) variable.

For the initial value problems (for which the IST method was originally developed), where the data and the solution are assumed to vanish sufficiently fast as the spatial variable approach infinities, the direct spectral problem takes the form of a scattering problem.
As for the inverse problem, the IST method in the original formulation uses the Gelfand--Levitan--Marchenko integral
equations \cite{novikov1984theory}. An alternative approach to the inverse problem is to consider a factorization problem of the Riemann--Hilbert
(RH) type, formulated in the complex plane of the spectral parameter involved in the Lax pair equations \cite{trogdon2015riemann}.

The extension of the IST method to problems formulated on the half-line or on an interval proved to be a challenging task. A systematic approach to these problems known as the Unified Transform Method (a.k.a. the Fokas Method) was introduced by Fokas \cite{Fokas1997}
and further developed by many researchers; see \cite{fokas2008unified} and references therein. The method is based on the simultaneous spectral analysis of both equations of the Lax pair and the subsequent analysis of the so-called Global Relation coupling, in the spectral terms, the appropriate spectral transforms of the initial data and all boundary values. In certain cases of boundary conditions (called \emph{linearizable}), the  Global Relation can be ``solved'' in a way that allows us to formulate the associated RH 
problem in terms of the data for a well-posed problem alone. 

In some recent papers \cite{DFL21, FL21}, it was shown  that the initial boundary value problem on a finite interval $[0,L]\subset\mathbb R$ with $x$-periodic boundary conditions 
[$q(0,t)=q(L,t)$, $q_x(0,t)=q_x(L,t)$] for the nonlinear Schr\"odinger (NLS) equation, the 
focusing NLS,
\begin{equation}\label{nls}
i q_t + q_{xx} + 2|q|^2 q = 0,
\end{equation}
as well as the defocusing NLS 
belong to the class of linearizable problems: 
the solution $q(x,t)$ can be expressed in terms of the solution of an RH problem, the data for which (the jump matrix across a certain contour and the residue conditions) can be expressed in terms of the entries of the scattering matrix for a spectral problem on the whole line, associated with $q(x,0)$, $x\in [0,L]$
(continued by $0$ on ${\mathbb R}\setminus [0,L]$). Particularly, the contour for the RH problem is a union of the real and the imaginary axes and a number (possibly infinite) of finite segments symmetric w.r.t. the real axis.

The NLS equation is known to be a 
key model governing under specific conditions the signal propagation in single-mode optical fibres \cite{Turitsyn:17, Essiambre2010}.
The linearization of problems for the NLS equation provided by the IST method was the basis for the development of the nonlinear Fourier transform (NFT) based
optical communication systems\cite{Turitsyn:17, Yousefi2014}. The central idea of such an approach is  to use, in order to carry the encoded data, the
so-called nonlinear spectrum of a signal
and to take advantage of the linear evolution of the spectrum. Most of the NFT-based
optical communication systems studied so far deal with the rapidly-vanishing signals and suffer from the burst mode operation and high computational complexity of the involved processing elements, which reduces the practicality of the approach \cite{Turitsyn:17, derevyanko2016capacity}.
Over recent years, there has been a growing interest in the development of optical
communication methods based on the utilization of non-decaying solutions to the NLS equation \cite{goossens2020data,kamalian2016periodic,kamalian2016periodic1,kamalian2018signal,JLT20,goossens2019experimental} as a more
efficient alternative to the ''conventional`` NFT-based communications \cite{le2015nonlinear}. The IST operations associated with periodic finite-genus
NLS solutions were named periodic nonlinear Fourier transform (PNFT) in these works. In Refs. \cite{kamalian2018signal,JLT20}, an optical signal
modulation and digital signal processing method has been proposed for a PNFT-based transmission, where the
inverse problem step (the constructing of a signal in the physical domain at the transmitter side) harnesses the numerical solution to the RH problem. 

The data for the RH
the problem associated with the NLS equation are $2\times 2$ jump matrices, which are off-diagonal matrices satisfying a certain symmetry condition with constant
entries on each separated part of the jump contour (consisting of a finite number of arcs). The solution of
such an RH problem gives rise to a finite-band (finite-genus) solution to the NLS equation 
\cite{belokolos1994algebro, AMSA17}: the endpoints of the arcs fix the associated Riemann surface, while the constants in the jump matrices specify an individual solution. 
The present paper mainly addresses the direct part of the approach: given the profile $q(x,T)$ associated with some fixed $t=T$, recover the constants in the 
jump matrices of the RH problem generating $q(x,T)$ via the solution of an RH problem having the form as described above. 
This is done through transforming the RH representation of periodic solutions developed in \cite{DFL21,FL21} to that involving the jumps across the ``bands'' alone. Since the latter RH problem can be solved explicitly in terms of the associated Riemann theta functions \cite{belokolos1994algebro, AMSA17}, we in particular obtain the affirmative answer to the hypothesis raised  in Remark 5.3 of \cite{FL21} that one can 
represent a theta-function finite-band solution in terms of the solution of a RH problem.
We note that the direct problem for the general quasi-periodic solutions was solved approximately using neural networks in Ref.~\cite{bogdanov2023phase}. 

The outline of the paper is as follows. In Section \ref{sec:inv}, we review the inverse part consisting of generating a finite-band solution to the NLS equation by solving the RH problem with appropriate data (jump conditions).
In Section \ref{sec:sketch}, we briefly describe
the ideas behind the development of the RH problem formalism for the periodic 
problem for the NLS equation 
presented in Refs.~\cite{DFL21} and \cite{FL21}. In Section \ref{sec:details}, we present the details of the sequence of transformations 
of the RH problem leading to our main results on the direct problem
stated in Theorems \ref{thm:step-2}--\ref{dir-main}. 
The evolution of the spectral data is discussed in Section \ref{sec:evolution}. Finally, in Section \ref{sec:examples}, we illustrate numerically the recovery of the phases (the constants in the jump matrices for the RH problem) 
using our direct problem algorithm.

\section{The inverse problem by the RH approach}\label{sec:inv}

A wide variety of solutions of an integrable nonlinear evolution equation can be constructed in terms of solutions to a family of RH problems (parameterized by the independent variables of the nonlinear equation, say, $x$ and $t$) whose data depends on $x$ and $t$ in a way specific to the integrable nonlinear equation in question. 
Specifically, in the case of the focusing NLS equation \eqref{nls},
we have the following.

\begin{prop}(\cite{belokolos1994algebro,AMSA17})\label{p2}
 Given $\{z_j\}_0^N$ with $\Im z_j>0$ and $\{\phi_j\}_0^N$ 
with $\phi_j\in [0,2\pi)$, define
\begin{itemize}
    \item 
    the oriented contour $\Sigma^N = \cup_{j=0}^N  \Sigma_j$, where 
		$\Sigma_j = (z_j, \bar z_j)$  (an arc connecting $z_j$ with $ \bar z_j$), and 
    \item 
    the $2\times 2$-valued function 
     \begin{equation}\label{J_j}
    J(x,t,z) = \begin{pmatrix}
    0 & ie^{-i\phi_j-2iz x-4iz^2 t} \\
  ie^{i\phi_j+2iz x+4iz^2 t} & 0
    \end{pmatrix}, \quad z\in\Sigma_j
    \end{equation}
\end{itemize}
\noindent and consider the following RH problem:
find 
a $2\times 2$-valued function $\Psi(x,t,z)$ such that: 
\begin{enumerate}
\item
For all $x\in \mathbb R$ and 
$t\in \mathbb R$,
$\Psi(x,t,z)$ is analytic w.r.t. $z$ for $z\in{\mathbb C}\setminus \bar\Sigma^N$ 
and continuous up to $\Sigma^N$ from the both sides of $\Sigma^N$;
\item
The limiting values $\Psi^+(x,t,z)$ and $\Psi^-(x,t,z)$, $z\in \Sigma^N$
of $\Psi(x,t,z)$, as $z$ approaches $\Sigma^N$ from the $+$ and $-$ side respectively,
 are related  by $J(x,t,z)$:
\begin{equation}\label{RHjump-Psi}
\Psi^+(x,t,z) = \Psi^-(x,t,z) J(x,t,z),\qquad z\in\Sigma^N;
\end{equation}
\item
At $z_j$ and $\bar z_j$,
$\Psi(x,t,z)$ has the inverse fourth root
singularities;
\item
As $z\to\infty$,
\begin{equation}\label{RHnorm-Psi}
\Psi(x,t,z) = I+O(1/z),\qquad I = 
\left(\begin{smallmatrix}
1 & 0 \\ 0 & 1 
\end{smallmatrix}\right).
\end{equation}


\end{enumerate}

Then
\begin{enumerate}
\item
 For all $x$ and $t$, the RH problem (\ref{J_j})--(\ref{RHnorm-Psi}) has a unique solutions 
 $\Psi(x,t,z)$, which satisfies 
 $\overline{\Psi(x,t,\bar z)}=\sigma_2 \Psi(x,t,z) \sigma_2$, where 
$\sigma_2 = \left(\begin{smallmatrix}
0 & -i \\ i & 0 
\end{smallmatrix}\right)$;
\item
 Defining $\Phi(x,t,z):= \Psi(x,t,z)e^{(-izx-2iz^2t)\sigma_3} $,
where $\sigma_3 = \left(\begin{smallmatrix}
1 & 0 \\ 0 & -1 
\end{smallmatrix}\right)
$,
determining $\Psi_1(x,t)$ from
$\Psi(x,t,z) = I+\frac{\Psi_1(x,t)}{z} + \dots$ as $z\to\infty$, and determining $q(x,t)$ by
\begin{equation}\label{q-RH}
q(x,t)=2i [\Psi_1]_{12}(x,t)
\end{equation}
(where $[\cdot]_{12}$ stands for the $12$ entry of a  matrix),
we have:
\begin{enumerate}
\item
$q(x,t)$ is a solution of \eqref{nls};
    \item 
    $\Phi(x,t,z)$ satisfies the system of linear differential equations (Lax pair) 
\begin{subequations}\label{Lax}
\begin{align}\label{Lax-x}
&\Phi_x = U\Phi \\
\label{Lax-t}
&\Phi_t = V\Phi
\end{align}
\end{subequations} 
with
\begin{subequations}\label{Lax-UV}
\begin{align}\label{Lax-U}
&U(x,t,z)= -iz\sigma_3 +\begin{pmatrix}
0 & q(x,t) \\
-\bar q(x,t) & 0 
\end{pmatrix},\\
\label{Lax-V}
&V(x,t,z)= -2iz^2\sigma_3 +2z \begin{pmatrix}
0 & q(x,t) \\
-\bar q(x,t) & 0 
\end{pmatrix} + \begin{pmatrix}
i|q|^2 & iq_x \\
i\bar q_x & -i|q|^2 
\end{pmatrix};
\end{align}
\end{subequations} 

\end{enumerate}
\item 
$q(x,t)$ given by \eqref{q-RH} 
is a solution of the NLS equation of finite-genus type: it can be expressed in terms 
of Riemann theta functions associated with the Riemann surface of genus $N$, with
the branch points at $z_j$ and $\bar z_j$, $j=0,\dots,N$.
    
\end{enumerate}

\end{prop}

The last statement of Proposition \ref{p2} follows from
the possibility to express $q(x,t)$ in terms of the solution
of another RH problem (see Proposition \ref{p3} below), which can be considered as a transformation
of the original RH problem evoking the so-called ''$g$-function mechanism`` \cite{DVZ94, DVZ98}.

In order to formulate the modified RH problem, we need a set of
parameters uniquely defined by the set of the branch points 
$z_j$ and $\bar z_j$, $j=0,\dots , N$. First, define $w(z)$
by
\begin{equation}\label{w}
    w(z) = \prod_{j=0}^N \sqrt{(z-z_j)(z-\bar z_j)}
\end{equation}
as a function analytic in ${\mathbb C}\setminus \Sigma^N$
whose branch is fixed by the asymptotic condition 
$w(z)\simeq z^{N+1}$ as $z\to\infty$. Let each arc $\Sigma_j$
be oriented upward and let $w^+(z)$ be the values of $w$ 
at the ``+'' side of the corresponding $\Sigma_j$.
Further, define the $N\times N$ matrix ${\mathbf K}$ by
\begin{equation}\label{K}
    {\mathbf K}_{mj}:=\int_{\Sigma_j}\dfrac{\xi^{m-1} d\xi}{w^+(\xi)}, \qquad m,j=1, \, \dots \, N,
\end{equation}
and determine the vectors ${\mathbf C}^f:=(C^f_1, \dots, C^f_N)^T$ and ${\mathbf C}^g:=(C^g_1, \dots, C^g_N)^T$ as
the solutions of the following linear equations:
\begin{equation}\label{CfCg}
    {\mathbf K}\cdot {\mathbf C}^f = \left[ 0,\dots,0,-2\pi i
    \right]^T, \qquad
    {\mathbf K}\cdot {\mathbf C}^g = -4\pi i
    \left[ 0,\dots,0,2, \sum_{j=0}^N (z_j+\bar z_j)
    \right]^T,
\end{equation}
(in the case $N=1$ or $N=2$, the last (respectively, two last) equations
from \eqref{CfCg} are to be considered).

Finally, determine the constants $f_0$ and $g_0$ from the large-$z$
developments of two scalar functions, $f(z)$ and $g(z)$, analytic in 
${\mathbb C}\setminus \Sigma^N$:
\begin{align}\label{f}
     & f(z)  :=  \frac{w(z)}{2\pi i}\sum_{j=1}^N \int_{\Sigma_j}
    \frac{C^f_j d\xi}{w^+(\xi)(\xi-z)} = z+f_0 +O(1/z), \\
    \label{g}
   &  g(z)  :=  \frac{w(z)}{2\pi i}\sum_{j=1}^N \int_{\Sigma_j}
    \frac{C^g_j d\xi}{w^+(\xi)(\xi-z)} = 2z^2+g_0 +O(1/z).
\end{align}

\begin{prop}(\cite{AMSA17})\label{p3}
Given $\{z_j\}_0^N$ with $\Im z_j>0$ and $\{\phi_j\}_0^N$ 
with $\phi_j\in [0,2\pi)$,
the genus-$N$ solution $q(x,t)$ of the NLS equation,
which can be obtained as the solution of the RH problem of Proposition \ref{p2},
can also be expressed by
\begin{equation}\label{q-Rhmod}
q(x,t) = 2i [\hat\Phi_1]_{12}(x,t) e^{2i f_0 x + 2i g_0 t},
\end{equation}
where $\hat\Phi_1$ enters the large-$z$ development
\begin{equation}
    \hat\Phi(x,t,z) = I + \frac{\hat\Phi_1(x,t)}{z} + \dots
\end{equation}
of the solution $\hat\Phi(x,t,z)$ of the following RH problem: find $\hat\Phi(x,t,z)$  analytic in 
${\mathbb C}\setminus \Sigma^N$ satisfying the jump conditions
\begin{equation}\label{hatphi-jump}
    \hat\Phi^+(x,t,z) = \hat\Phi^-(x,t,z)\hat J_j(x,t),\qquad z\in \Sigma_j, \ j=0,\dots, N
\end{equation}
with
\begin{equation}\label{hatj}
    \hat J_j(x,t) = \begin{pmatrix}
    0 & ie^{-i(\phi_j+C^f_j x + C^g_j t)} \\
    ie^{i(\phi_j+C^f_j x + C^g_j t)} & 0
    \end{pmatrix}
\end{equation}
and the normalization condition
\begin{equation}\label{hatphi-norm}
    \hat\Phi(x,t,z) = I +O(1/z),\qquad z\to\infty.
\end{equation}
Here $C^f_0 = C^g_0=0$ whereas the constants $f_0$ and $g_0$ in (\ref{q-Rhmod})
and $C^f_j$, $C^g_j$, $j=1,\dots,N$ in (\ref{hatj})
are determined by  $\{z_j\}_0^N$ 
via (\ref{K})--(\ref{g}).
\end{prop}
 
\begin{rem}
The solution $\Psi(x,t,z)$ of the RH problem in Proposition \ref{p2}  is related to the solution $\hat\Phi(x,t,z)$ of the RH problem (\ref{hatphi-jump})--(\ref{hatphi-norm}) as follows:
\begin{equation}\label{RH-rel}
    \Psi(x,t,z) = e^{(if_0 x+ig_0 t)\sigma_3}
    \hat\Phi(x,t,z) e^{(i(z-f(z)) x+i(2z^2-g(z)) t)\sigma_3}.
\end{equation}

\end{rem}
\begin{rem}
 It is the RH problem (\ref{hatphi-jump})--(\ref{hatphi-norm})
 that can be solved explicitly \cite{belokolos1994algebro,AMSA17}, in terms of Riemann theta functions 
 associated with the genus-$N$ Riemann surface
 associated with $w(z)$ (\ref{w}) and characterized by the branch points  $z_j$ and 
$\bar z_j$, $j=0,\dots, N$.
 \end{rem}
 \begin{rem}
If all $C^f_j$ together with $f_0$
 turn to be commensurable, then the underlying solution
 of the NLS equation is periodic in $x$. 
 \end{rem}

\section{Direct problem in the periodic case: a sketch}\label{sec:sketch}
 
 The direct problem associated with the RH problem \eqref{J_j}--\eqref{RHnorm-Psi}
 (i.e., with the problem: given $\{\phi_j\}_0^N$, construct $q(x,t)$) consists in the following:
 given a $N$-genus solution $q(x,t)$ of the NLS equation associated with the prescribed branch points $\{{z}_j\}_0^N$ and 
 evaluated as a function of $x$ at some fixed $t=t^*$, 
determine the underlying phase parameters $\phi_j$. 

In the case where $C^f_j$ together with $f_0$ (see (\ref{CfCg})--(\ref{g}))
are commensurable  and thus the underlying solution of the NLS equation is periodic in $x$, a possible way to solve the direct problem is based on the idea of finding a RH representation for the solution of the initial boundary value problem (IBVP) for the NLS equation, where the initial data given  for $x$ varying on an interval (of the periodicity length $L$), i.e., $q(x,0) = q_0(x)$ for $x\in (0,L)$, are supplemented by the periodicity conditions:
\begin{equation}\label{period}
    q(0,t)=q(L,t),\quad q_x(0,t)=q_x(L,t)\quad \text{for all}\  t\ge 0.
\end{equation}
If the RH problem in this representation had  the same structure as the original RH problem \eqref{J_j}--\eqref{RHnorm-Psi},  then the constants $\phi_j$ appearing in the jump construction would give the sought solution of our direct problem.

To get the appropriate representation, one can proceed in two steps: (i) first, provide \emph{some} RH representation (with \emph{some} contour and jumps), where the data for the RH problem can be constructed from the data of the periodic IBVP, i.e., the initial data $q_0(x)$ for $x\in (0,L)$;
(ii) second, using the flexibility of the RH
representation for the solution of nonlinear equations, transform this (original) RH problem to that having the above-mentioned \emph{desired form} \eqref{J_j}.

The first step has been recently addressed in \cite{DFL21} and \cite{FL21}, where it was shown that in the case (in particular)  of the focusing NLS equation, the solution of the periodic IBVP 
(not necessarily finite-band)
can be given in terms of the solution of a RH problem, where (i) the contour is the union of a (possibly infinite) number of finite arcs and the real and imaginary axes, and (ii) the jump
matrices can be constructed in terms of the entries $a({z})$ and $b({z})$ of the scattering matrix:
\begin{equation}\label{s}
    s({z})=\begin{pmatrix}
\overline{a(\bar z)} & b({z}) \\ -\overline{b(\bar z)} & a({z})
\end{pmatrix} \equiv \begin{pmatrix}
a^*(z) & b({z}) \\ -b^*(z) & a({z})
\end{pmatrix},
\end{equation}
where we adopt the notation $a^*(z) = \overline{a (\bar z)}$ etc.
Here $s(z)$ is the scattering matrix of the Zakharov--Shabat spectral problem (the $x$-equation of the Lax pair for the NLS equation) \eqref{Lax-x}
considered on the whole line,
with the potential $q=q(x,0)$ involved in $U$ being 
continued on the whole line by setting it to $0$ for $x$ outside $[0,L] $. 

To ensure the consistency of presentation, we briefly describe this step that 
can be performed in two sub-steps.
In sub-step 1, an RH problem is constructed
using the spectral functions $a({z})$ and $b({z})$ 
supplemented by the  spectral functions $A({z})$, $B({z})$, 
$A_1({z})$, $B_1({z})$, 
that enter the scattering matrices
\[S({z})=\begin{pmatrix}
A^*(z) & B({z}) \\ -B^*(z) & A({z})
\end{pmatrix},\qquad 
S_1({z})=\begin{pmatrix}
A_1^*(z) & B_1({z}) \\ -B_1^*(z) & A_1({z})
\end{pmatrix}
\]
associated with the 
$t$-equation from the Lax pair \eqref{Lax-t} considered for $x=0$ and $x=L$
respectively
\cite{fokas2004nonlinear, FIS}.

Namely, assuming for a moment that $q(0,t)$
and $q_x(0,t)$ are given for  $t\in(0,T)$ with  some $T>0$,
Eq. (\ref{Lax-t}) can be considered,  similarly to (\ref{Lax-x}), as a spectral problem for a matrix equation with coefficients determined in terms of $q(0,t)$ and $q_x(0,t)$, giving rise to $S(z)$ as the 
associated scattering matrix. Similarly, 
$q(L,t)$ and $q_x(L,t)$ give rise to $S_1(z)$.
Then the periodicity condition \eqref{period} implies that $S(z)=S_1(z)$.
Since $V$ in Eq.~(\ref{Lax-t}) is a polynomial of the second order w.r.t. ${z}$, it follows that the contour where the scattering relation is established consists of two lines, the real and imaginary axes (where $\Im {z}^2=0$). 

Since neither $q(0,t)$ nor $q_x(0,t)$ 
are given as the data for the periodic IBVP,  sub-step 2 addresses the problem of replacing the RH problem constructed in terms of $a({z})$, $b({z})$, $A({z})$, and $B({z})$ by an equivalent one (in the sense that $q(x,t)$ obtained following (\ref{q-RH}) from the 
both problems are the same), whose formulation involves  $a({z})$ and $b({z})$ only. A key for performing this sub-step is the so-called Global Relation~\cite{fokas2008unified, DFL21, fokas2004nonlinear, FIS}, which is a relation amongst 
$a({z})$, $b({z})$, $A({z})$, and $B({z})$ reflecting the fact that the IBVP with periodic boundary conditions is well-posed (particularly, has a unique solution) without prescribing the boundary values $q(0,t)$ and $q_x(0,t)$.

In the current setting (i.e., for the periodic problem in $x$), the global relation takes the form of the equation:
\begin{align}\label{GR1}
    e^{2i{z} L}\left(A({z})a^*({z})+B({z})b^*({z})\right)B({z}) +\left(A({z})b({z})-a({z})B({z})\right)A({z}) =\\ \nonumber
    = e^{4i{z}^2 T} O\left(
    \frac{1+e^{2i{z} L}}{{z}}
    \right),
\end{align}
where the r.h.s. is not given precisely but only asymptotically, as ${z}\to\infty$. Noticing that the r.h.s. in (\ref{GR1}) approaches $0$ as ${z}\to\infty$ staying in the first quadrant of the complex ${z}$-plane suggests replacing the r.h.s. by zero, which leads to a quadratic equation for the ratio $B({z})/A({z})$, with the coefficients given in terms of $a({z})$ and $b({z})$. Define $R({z})$ as the solution of the 
resulting equation,
\begin{equation}\label{R-equation}
e^{2izL}b^*(z)R^2(z)
+ \left(e^{2izL} a^*(z) - a(z)\right) R(z)  +b(z) = 0,
\end{equation}
by
\begin{equation}\label{R}
    R({z}) = \frac{e^{-i{z} L} a({z})-e^{i{z} L} a^*(z)
    +\sqrt{(e^{-i{z} L} a({z})-e^{i{z} L} a^*(z))^2-4b^*(z)b({z})}}{2e^{i{z} L}b^*(z)},
\end{equation}
where the branch of the square root is chosen such that the branch cuts are the arcs connecting the pairs of complex conjugate points (actually, they are ${z}_j$ and 
$\bar z_j$) and that $R({z})\to 0$ as ${z}\to\infty$. Then, one can show that the RH problem sought in sub-step 2 is that obtained from the original RH problem, where $B({z})/A({z})$ is replaced by $R({z})$. Due to the jumps of $R({z})$ across the arcs connecting ${z}_j$ and $\bar z_j$,  additional jump conditions on these arcs arise and thus the jump contour takes the form: $\cup_{j=0}^N  \Sigma_j \cup {\mathbb R}\cup i{\mathbb R}$,
whereas the jump matrix on all parts of the contour 
can be algebraically given in terms of 
$a({z})$, $b({z})$, and $R({z})$. To complete the formulation of the RH problem from step 1, the jump conditions have to be complemented by the residue conditions at the singularities of $R({z})$, if any (these are also given in terms of spectral quantities determined by the initial data only). For the exact formulation of the RH problem of step 1, see \cite{FL21}, Theorem 4.6\footnote{In \cite{FL21},
the notation $\tilde \Gamma$ is adopted for $R$.} and Theorem \ref{RHP} below.

\textbf{Assumptions.} In order to fix ideas while avoiding technicalities,
we assume that (i) $a(z)$ has a finite number of simple zeros  in the upper complex  half-plane
 and  these zeros do not coincide with the poles of $R({z})$ and 
$ R^*({z})$ and (ii)
 $\Re z_j\ne 0$ for all $j$.

The second step consists of transforming the RH problem 
described above (with jumps across $\cup_{j=0}^N  \Sigma_j \cup {\mathbb R}\cup i{\mathbb R}$ and residue conditions)
to a RH problem of the form \eqref{J_j}--\eqref{RHnorm-Psi}
 with some constants $\phi_j$. We will show that this step can also be divided into 
 several sub-steps:
(i) transforming the RH problem to that with jumps across $\mathbb R$ and $i\mathbb R$
having the diagonal structure;
(ii) reducing the jump conditions to those 
across $ \Sigma^N = \cup_{j=0}^N  \Sigma_j$ only and 
getting rid of singularity conditions;
(iii) making the jumps on each $\Sigma_j$ to have the structure 
 as in (\ref{J_j}).

In the  case $N=0$, this step has been done in \cite{DFL21,FL21}; 
in this case, the contour for the RH problem consists 
of a single arc, and there are no singularity conditions. The associated ($0$-genus) solution of the NLS equation is a simple exponential function:
$q(x,t)=\alpha e^{-2i\beta x +2i\omega t +i\phi_0}$,
where $\alpha=\Im {z}_0$, $\beta=\Re {z}_0$, and 
$\omega=\alpha^2-2\beta^2$.

The cases with $N\ge 1$ turn out to be more involved.
Particularly, in the realization of  sub-step (ii)
we need to get rid of singularity conditions at the singularity points of $R({z})$. In terms of the spectral theory of the Zakharov--Shabat equation with 
periodic coefficients, the (possibly empty) set of such singularity points $\{\mu_j\}_1^{N_1}$, $N_1\le N$ consists of those conjugated
\emph{auxiliary spectrum} points for this problem
which are located on the sheet (of the two-sheeted
Riemann surface of $R$) characterized by the condition
$R({z})\to 0$ as ${z}\to\infty$.

 The resulting (${z}$-dependent) jump matrix is as follows:
\begin{equation}\label{rhp-check}
\check{J}(x,t,{z})=\begin{pmatrix}
0 & iJ_{00}({z})e^{-2i{z} x-4i{z}^2 t }\\
iJ_{00}^{-1}({z})e^{2i{z} x+4i{z}^2 t} & 0
\end{pmatrix},\qquad {z}\in \Sigma^N,
\end{equation}
where 
$J_{00}({z})$ can be expressed in terms of $R(z)$ (see \eqref{P-poles} below). 

Having $J_{00}({z})$ obtained, sub-step (iii)
 can be done using the solution
of the scalar RH problem:
\begin{align}\label{d}
& d_+({z}) d_-({z})=J_{00}({z}) e^{i\phi_j},\quad {z}\in \Sigma_j,\quad j=0,\, \dots, \, N,\\ \label{d1}
& d({z})\to 1, \quad {z}\to\infty.
\end{align}
In this problem, the constants $\phi_j$ are not prescribed  but determined uniquely 
by \eqref{d1} applied to the Cauchy-type solution of 
(\ref{d}); they are the phases sought in the direct problem.

\section{Direct problem in the periodic case: details}\label{sec:details}

As we have mentioned above, using the ideas of the Unified Transform Method, 
 it is possible to represent the solution of the periodic problem
 \begin{subequations}\label{periodic-NLS}
	\begin{equation}\label{nls-ibvp}
i q_t + q_{xx} + 2|q|^2 q = 0, \qquad x\in (0,L), \ t>0;
\end{equation}
\begin{equation}\label{ini-ibvp}
q(x,0) = q_0(x), \qquad x\in [0,L];
\end{equation}
	\begin{equation}\label{period-ibvp}
    q(0,t)=q(L,t),\quad q_x(0,t)=q_x(L,t)\quad \text{for all}\  \, t\ge 0,
\end{equation}
\end{subequations}
in terms of the solution of an RH problem, the data for which 
(jump and residue conditions) can be constructed using the spectral functions $a(z)$
and $b(z)$ uniquely determined by the initial data $q_0(x)$. 
Namely, $a(z)$
and $b(z)$ are the entries of the scattering matrix $s(z)$ \eqref{s} relating the dedicated 
solutions $\Phi^0_2(x,z)$ and $\Phi^0_3(x,z)$ of the Zakharov--Shabat equation \eqref{Lax-x},
\eqref{Lax-U} taken at $t=0$: let $U_1(x,0):=\begin{pmatrix}
    0 & q_0(x) \\-\bar q_0(x) & 0
\end{pmatrix}$; then $s(z)$ is determined by
\[
\Phi^0_3(x,z) = \Phi^0_2(x,z) s(z),
\]
where $\Phi^0_2(x,z)$ and $\Phi^0_3(x,z)$ 
are the  solutions of the integral equations
\[
\begin{aligned}
    \Phi^0_2(x,z) & = e^{-izx\sigma_3}+\int_0^x e^{-iz(x-y)\sigma_3}U_1(y,0)\Phi^0_2(y,z)dy,\\
    \Phi^0_3(x,z) & = e^{-izx\sigma_3}-\int_x^L e^{-iz(x-y)\sigma_3}U_1(y,0)\Phi^0_3(y,z)dy.
\end{aligned}
\]

In the construction of the associated RH problem,
a key role is played by $R(z)$ \eqref{R}. Before presenting this RH problem, we discuss some analytic properties of $R(z)$.

\subsection{Analytic properties of $R(z)$}

The scattering matrix $s({z})$ in our setting is closely related to the 
monodromy matrix $\mathcal{M}({z})$ of the Zakharov--Shabat equation with periodic conditions defined as $\mathcal{M}({z})=\Phi(L,0,{z})$, where $\Phi(x,0,{z})$ is the solution of Eq.~(\ref{Lax-x}) satisfying the condition $\Phi (0,0,{z})=I$.
Particularly, we have 
\begin{equation}\label{M-sym}
\mathcal{M}_{11}(z) = \mathcal{M}^*_{22}(z) =  e^{-izL}a(z), \quad \mathcal{M}_{12}(z) = - \mathcal{M}^*_{21}(z) = -e^{-izL}b(z).
\end{equation}

In terms of $\mathcal{M}_{ij}$, equation \eqref{R-equation} reads as 
\begin{equation}\label{R-eq-2}
\mathcal{M}_{21}(z) R^2(z) +(\mathcal{M}_{22}(z)-\mathcal{M}_{11}(z))R(z) - \mathcal{M}_{12}(z) = 0;
\end{equation}
its solutions,
$R_{(1)}({z})$ and $R_{(2)}({z})$, can be expressed as follows:
\begin{subequations}\label{R-1-2-M}
\begin{align}\label{R-1-M}
   & R_{(1)}({z})=\frac{\mathcal{M}_{11}({z})-\mathcal{M}_{22}({z})
    -\sqrt{\Delta^2({z})-4}}{2\mathcal{M}_{21}({z})} = 
    \frac{e^{-izL}a(z)-e^{izL}a^*(z)
    -\sqrt{\Delta^2({z})-4}}{e^{izL}b^*(z)}, \\
    \label{R-2-M}
	&	R_{(2)}({z})=\frac{\mathcal{M}_{11}({z})-\mathcal{M}_{22}({z})
    +\sqrt{\Delta^2({z})-4}}{2\mathcal{M}_{21}({z})}  = 
    \frac{e^{-izL}a(z)-e^{izL}a^*(z)
    +\sqrt{\Delta^2({z})-4}}{e^{izL}b^*(z)}, 
\end{align}
\end{subequations}
where 
\begin{equation}\label{Del}
    \Delta ({z}) :=\mathcal{M}_{11}({z}) + \mathcal{M}_{22}({z}) 
= e^{-izL}a(z) + e^{izL}a^*(z),
\end{equation}
and we have used that 
\begin{equation}\label{M-det}
\det \mathcal{M}(z) =  \mathcal{M}_{11}(z)\mathcal{M}_{22}(z) - \mathcal{M}_{12}(z)\mathcal{M}_{21}(z) = 
a^*(z)a(z)+b^*(z)b(z)\equiv 1.
\end{equation}

As functions of $z$, $R_{(1)}({z})$ and $R_{(2)}({z})$ can be viewed
as the branches of function $R$ meromorphic on the Riemann surface $(z,w)$
of 
\[
w^2 = \prod\limits_{j=0}^N (z-z_j)(z-\bar z_j)
\]
 \emph{assuming} that there is a finite number (denoted
by $N+1$) of conjugated pairs $\{z_j,\bar z_j\}$ of \emph{simple zeros} of function $\Delta^2(z)-4$.

In the context of the spectral theory of the Zakharov--Shabat equation with periodic conditions, $\{{z}_j, \bar z_j\}_0^N$  are 
called the \emph{main spectrum}; they are 
the branch points of $R$.
On the other hand, the simple zeros of $M_{12}({z})$
which are not double zeros of $\Delta^2({z})-4$
(as well as the multiple zeros of $\mathcal{M}_{12}({z})$)
constitute the \emph{auxiliary spectrum}
$\{\mu_j\}_1^{N}$.

Notice that by the definition of $\mathcal{M}$,
all zeros of $\mathcal{M}_{12}(z)$ are  the eigenvalues of the homogeneous Dirichlet-type problem 
for the Zakharov--Shabat equation \eqref{Lax-x} on $(0,L)$ with $q=q_0(x)$: 
if $\mathcal{M}_{12}(\tilde z) = 0$ for some $\tilde z$, then there exists a non-trivial 
vector solution $\Xi(x,\tilde z)=(\Xi_1(x,\tilde z), \Xi_2(x,\tilde z)^T$ of \eqref{Lax-x} such that 
$\Xi_1(0,\tilde z)=\Xi_1(L,\tilde z)=0$ (actually, one can take
$\Xi(x,\tilde z) = \Phi^{(2)}(x,\tilde z)$, where ${M}^{(l)}$ denote the $l$-th column of a $2\times 2 $ matrix $M$).

Similarly, all zeros of $\mathcal{M}_{21}(z)$ are 
the eigenvalues of the homogeneous Neumann-type problem 
for the Zakharov--Shabat equation \eqref{Lax-x} on $(0,L)$: for such $z$ there exists a 
non-trivial 
vector solution  of \eqref{Lax-x} such that
its second component equals $0$ at $x=0$
and $x=L$.

One can view $R_{(1)}({z})$ and $R_{(2)}({z})$ 
as meromorphic functions on ${\mathbb C}\setminus \Sigma^N$ with the branch cut 
$\Sigma^N$, where 
$\Sigma^N=\cup_0^N \Sigma_j$ and $\Sigma_j$ are the vertical segment connecting 
$z_j$ and $\bar z_j$. Particularly, we specify $R_{(1)}({z})$ by the condition 
$R_{(1)}({z})\to 0$ as $z\to\infty$.

Let's list some analytic properties of $R$ that hold for all $z$ 
including the limiting values at each side, $z_+$ and $z_-$, of $\Sigma_j(z)$:

\begin{enumerate}
	\item 
	By the definition of $R_{(1)}$ and $R_{(2)}$ (as the solutions of the quadratic equation),
	\begin{equation}	\label{R-prod}
	R_{(1)}({z}) \cdot R_{(2)}({z}) = -\frac{\mathcal{M}_{12}(z)}{\mathcal{M}_{21}(z)}
	= \frac{e^{-izL}b(z)}{e^{izL}b^*(z)};
\end{equation}
\begin{equation}	\label{R-sum}
	R_{(1)}(z) +  R_{(2)}(z) = \frac{\mathcal{M}_{11}(z)- \mathcal{M}_{22}(z)}{\mathcal{M}_{21}(z)}
	=\frac{e^{-izL}a(z)-e^{izL}a^*(z)}{e^{izL}b^*(z)};
\end{equation}
\item
By the symmetries \eqref{M-sym},
\begin{equation}	\label{R_1-R_2}
	R_{(2)}^*(z) = -\frac{1}{R_{(1)}(z)},
\end{equation}
(the both sides of \eqref{R_1-R_2} satisfy the same quadratic equation)
and thus, in view of \eqref{R-prod},
\begin{equation}	\label{R-R-star-1}
	\mathcal{M}_{21}(z)R(z) = \mathcal{M}_{12}(z)R^*(z)
\end{equation}
or 
\begin{equation}	\label{R-R-star}
	e^{izL}b^*(z)R(z) = -e^{-izL}b(z)R^*(z),
\end{equation}
where $R(z) = R_{(1)}(z)$ or $R(z) = R_{(2)}(z)$.
\item
\begin{equation}\label{a-b-R}
\left(a(z)+b(z)R^*(z)\right)\left(a^*(z)+b^*(z)R(z)\right)\equiv 1.
\end{equation}

Indeed, 
\[
\begin{aligned}
\left(a +bR^*\right)&\left(a^*+b^*R\right)  = 
(\mathcal{M}_{11}-\mathcal{M}_{12}R^*)(\mathcal{M}_{22}+\mathcal{M}_{21}R) 
 = (\mathcal{M}_{11}-\mathcal{M}_{21}R) \\& \times (\mathcal{M}_{22}+\mathcal{M}_{21}R) 
 = 
1+\mathcal{M}_{12}\mathcal{M}_{21} -\mathcal{M}_{21}\left(\mathcal{M}_{21}R^2 +(\mathcal{M}_{22}-\mathcal{M}_{11})R\right) \\
& = 
1 - \mathcal{M}_{21}\left(\mathcal{M}_{21}R^2 +(\mathcal{M}_{22}-\mathcal{M}_{11})R-\mathcal{M}_{12}\right) =  1,
\end{aligned}
\]
where we have used \eqref{R-eq-2}, \eqref{M-det} and  \eqref{R-R-star}.
\item
\begin{equation}\label{ne0}
    a(z)+b(z)R^*(z) \ne 0
\end{equation}
for all $z$. This follows  from \eqref{a-b-R} and the fact that
$b^*(z)R(z)$ is, by \eqref{R-1-2-M}, non-singular.

Actually, this also follows from the representation:
\[
a^*(z)+b^*(z) R(z) = \frac{1}{2}e^{-izL}(\Delta(z) \pm \sqrt{\Delta^2(z)-4}).
\]
 \end{enumerate}

Finally,  we list some  properties of $R$ involving
 the limiting values at the different sides of $\Sigma_j(z)$
 (denoting $R^\pm(z):=R(z_\pm)$):
 
 \begin{enumerate}
 \item
 $\frac{R^+(z)}{R^{*+}(z)} = \frac{R^-(z)}{R^{*-}(z)}$
 (follows from \eqref{R-R-star-1} and the fact that $M_{ij}(z)$ are entire functions).
 \item
 \begin{equation}\label{R-pm}
 R^{*+}(z)R^-(z) = R^+(z)R^{*-}(z) = -1
 \end{equation}
 (follows from \eqref{R_1-R_2} and ${R_{(2)}}(z_+) = {R_{(1)}}(z_-)$,
 ${R_{(1)}}(z_+) = {R_{(2)}}(z_-)$);
  \item
  \begin{equation}\label{R-pm-a-b}
  a^*(z)+b^*(z)R^{-}(z) =  e^{-2izL}\bigl(a(z)+b(z)R^{*+}(z)\bigr)
 \end{equation}
 (follows from \eqref{R-pm}, \eqref{M-sym} and \eqref{R-eq-2} for $R^-$).
 
 \end{enumerate}

\subsection{RH problem associated with the periodic problem for the NLS}

From now on, we 
denote by $R(z)$ the branch in \eqref{R-1-2-M} decaying to $0$ as $z\to \infty$.
Define  $G(z)$, $G_1(z)$, and $G_2(z)$ 
 as follows:
 \begin{align}\label{G-R-1}
& G(z) =-\frac{R^*(z)}{a(z)(a(z)+b(z)R^*(z))}
  = -\frac{R^*(z)}{a(z)}(a^*(z)+b^*(z)R(z)) = -e^{-2izL}R^*(z)-\frac{b^*(z)}{a(z)}, \\
	\label{G-R-2}
& G_1(z) = \frac{e^{2izL}a(z)R(z)}{a(z)+b(z)R^*(z)}
= e^{2izL}a(z)R(z)(a^*(z)+b^*(z)R(z))  = a^2(z)\left(R(z)-\frac{b(z)}{a(z)}\right), \\
	\label{G-R-3}
& G_2(z) = a^{-2}(z)  G_1(z)   = R(z)-\frac{b(z)}{a(z)}.
 \end{align}
Using these functions, define a $2\times 2$ function $J_0(z)$ for 
$z\in {\mathbb R}\cup i{\mathbb R}\cup\Sigma^N$:
\begin{equation}\label{J0-2}
	J_0(z) = \begin{cases}
	\begin{pmatrix}
	1 & 0 \\ G_2^*(z) & 1
	\end{pmatrix}
	\begin{pmatrix}
	1 & \tilde r(z) \\ \tilde r^*(z) & 1+|\tilde r|^2(z)
	\end{pmatrix}
	\begin{pmatrix}
	1  & G_2(z) \\ 0 & 1
	\end{pmatrix}, & z\in {\mathbb R}_+,\\
	\begin{pmatrix}
	1 & G^*(z) \\ 0 & 1
	\end{pmatrix}
	\begin{pmatrix}
	1+|r|^2(z) & r^*(z) \\ r(z) & 1
	\end{pmatrix}
	\begin{pmatrix}
	1  & 0 \\ G(z)  & 1
	\end{pmatrix}, & z\in {\mathbb R}_-,\\
	\begin{pmatrix}
	1  & -G_2(z) \\ 0 & 1
	\end{pmatrix}
 \begin{pmatrix}
            a^{-1}(z) & 0 \\ 0 & a(z)
        \end{pmatrix}
	\begin{pmatrix}
	1  & 0 \\ G(z)  & 1
	\end{pmatrix}
 , & z\in i{\mathbb R}_+,\\
	\begin{pmatrix}
	1  & G^*(z) \\ 0 & 1
	\end{pmatrix}
 \begin{pmatrix}
            a^*(z) & 0 \\ 0 & (a^*(z))^{-1}
        \end{pmatrix}
	\begin{pmatrix}
	1  & 0 \\ -G_2^*(z)  & 1
	\end{pmatrix}, & z\in i{\mathbb R}_-,\\
	\begin{pmatrix}
	1  & R^+(z)-R^-(z) \\ 0 & 1
	\end{pmatrix}, & z\in \Sigma^N\cap I, \\
	\begin{pmatrix}
	1  & 0 \\
	e^{-2izL}({R^*}^-(z)-{R^*}^+(z))  & 1
	\end{pmatrix}, & z\in \Sigma^N\cap II, \\
	\begin{pmatrix}
	1  & 
	e^{2izL}({R}^+(z)-{R}^-(z)) \\ 0  & 1
	\end{pmatrix}, & z\in \Sigma^N\cap III, \\
	\begin{pmatrix}
	1  & 0 \\
	{R^*}^-(z)-{R^*}^+(z)  & 1
	\end{pmatrix}, & z\in \Sigma^N\cap IV,
	\end{cases}
	\end{equation}
 where 
 \begin{equation}
 r(z)=\frac{b^*(z)}{a(z)}, \qquad 
     \tilde r(z)=\frac{b(z)}{a(z)}.
 \end{equation}
Finally, specify the residue conditions for a $2\times 2$ function $M(x,t,z)$
at the poles of $R(z)$ and $R^*(z)$ as follows: 
 \begin{enumerate}
    \item 
    At the poles $\xi_j$ of $R(z)$ for $z\in I$:
    \begin{equation}\label{M-R-res-3}
		\underset{z=\xi_j}{\operatorname{Res}} M^{(2)}(x,t,z) = e^{-2i\xi_j x-4i\xi_j^2 t}
		\underset{z=\xi_j}{\operatorname{Res}}R(z) M^{(1)}(x,t,\xi_j).
		\end{equation}
	\item
 At the poles $\xi_j$ of $R^*(z)$ for $z\in II$:
	\begin{equation}\label{M-R-res-4}
		\underset{z=\xi_j}{\operatorname{Res}} M^{(1)}(x,t,z) = -e^{2i\xi_j (x-L)+4i\xi_j^2 t}
		\underset{z=\xi_j}{\operatorname{Res}}R^*(z) M^{(2)}(x,t,\xi_j).
		\end{equation}
    \item 
    At the poles $\xi_j$ of $R(z)$ for $z\in III$:
    \begin{equation}\label{M-R-res-5}
		\underset{z=\xi_j}{\operatorname{Res}} M^{(2)}(x,t,z) = e^{-2i\xi_j (x-L)-4i\xi_j^2 t}
		\underset{z=\xi_j}{\operatorname{Res}}R(z) M^{(1)}(x,t,\xi_j).
		\end{equation}
  \item
 At the poles $\xi_j$ of $R^*(z)$ for $z\in IV$:
	\begin{equation}\label{M-R-res-6}
		\underset{z=\xi_j}{\operatorname{Res}} M^{(1)}(x,t,z) = -e^{2i\xi_j x+4i\xi_j^2 t}
		\underset{z=\xi_j}{\operatorname{Res}}R^*(z) M^{(2)}(x,t,\xi_j).
		\end{equation}
\end{enumerate}

\begin{theorem}\label{RHP}
Let $a(z)$ and $b(z)$ be the spectral functions
associated with $q_0(x)$, $x\in (0,L)$ via
the solution of the direct scattering problem for 
the Zakharov--Shabat equation \eqref{Lax-x} 
with $q=q_0(x)$. Assume that 
(i) $a(z)$ has a finite number of simple zeros in ${\mathbb C}_+$ and 
(ii) the number of pairs $\{z_j,\bar z_j\}$ of simple zeros of
the function $\Delta^2(z)-4$, where $\Delta(z)$
is defined by \eqref{Del}, is finite. 
Introduce $\Sigma^N$ by  $\Sigma^N=\cup_0^N \Sigma_j$, where  $\Sigma_j$ is the vertical segment connecting 
$z_j$ and $\bar z_j$.
Let $R(z)$ be determined by $a$ and $b$ via \eqref{R-1-2-M} such that $R(z)$ is analytic in ${\mathbb C}\setminus \Sigma^N$ and $R(z)\to 0 $ as $z\to \infty$,
and let  $G(z)$ and $G_2(z)$ be determined in terms of $a$, $b$, and $R$ by \eqref{G-R-1} and \eqref{G-R-2}.

Let $q(x,t)$ be defined by 
$q(x,t) = 2i [M_1]_{12}(x,t)$, where 
$M(x,t,z)=I+\frac{M_1(x,t)}{z} + \dots$ as $z\to\infty$
and $M(x,t,z)$ is the solution of the Riemann--Hilbert
problem specified by (i) the jump conditions 
	\begin{equation}	\label{M-jump-1}
	M^+(x,t,z) = M^-(x,t,z)J(x,t,z), \qquad z\in {\mathbb R}\cup i{\mathbb R}\cup\Sigma^N,
	\end{equation}
	where contour is oriented such that ${\mathbb R}$ is oriented
	from left to right,  $i{\mathbb R}_+$  and $i{\mathbb R}_-$
	are oriented towards infinities, and $\Sigma_j$ are oriented upwards,
	from $\bar z_j$ to $z_j$,
 and
	\begin{equation}	\label{J-RHP}
	J(x,t,z) = e^{-(izx + 2iz^2 t)\sigma_3}J_0(z) e^{(izx + 2iz^2 t)\sigma_3},
	\end{equation}
where $J_0$ is given by 
\eqref{J0-2}; (ii) the residue conditions 
\eqref{M-R-res-3}--\eqref{M-R-res-6}, 
and (iii) the normalization condition $M(x,t,z)\to I $
as $z\to\infty$. Then $q(x,t)$ is the solution of the periodic problem \eqref{periodic-NLS}.
\end{theorem}

\begin{rem}
    As we mentioned above, the construction of the RH problem in Theorem \ref{RHP}
    is motivated by the application of the Unified Transform Method to the periodic problem
    \cite{DFL21,FL21}. On the other hand, one can show \emph{directly} that $q(x,t)$ obtained
    via the solution of this RH problem solves problem \eqref{periodic-NLS}.
\end{rem}
Particularly, one can prove that 
 $q(x,t)$ (i) satisfies the initial conditions $q(x,0)=q_0(x)$ and (ii) 
satisfies the periodicity conditions \eqref{period-ibvp} 
by proving that 
 (a) the jumps and the residue conditions for $t=0$ can be mapped 
to those in the RH problem associated with $q_0(x)$ and (b) the jumps and the residue conditions
for $x=0$ and $x=L$ can be mapped to each other.

\subsubsection{Verifying the initial conditions}

Recall that the RH problem associated with $q_0(x)$ is as follows
\cite{fokas2004nonlinear}: find $M^{(x)}(x,z)$ such that 
\begin{enumerate}
\item
$M^{(x)}(x,z)$ is meromorphic in ${\mathbb C}\setminus {\mathbb R}$
and satisfies the jump condition on $\Sigma^{(x)}:={\mathbb R}$:
\begin{equation}	\label{M-x-jump}
	M^{(x)+}(x,z) = M^{(x)-}(x,z)J^{(x)}(x,z), \qquad z\in \Sigma^{(x)},
	\end{equation}
where 
\[
	J^{(x)}(x,z) := e^{-izx\sigma_3}J_0^{(x)}(z) e^{izx\sigma_3},
	\]
	with
\begin{equation}\label{J0-x}
	J_0^{(x)}(z) = 
	\begin{pmatrix}
	1+|r|^2(z) & r^*(z) \\ r(z) & 1
	\end{pmatrix}.
\end{equation}	
\item
Assuming that $a(z)$ has a finite number of simple zeros $\{\nu_j\}_1^Q$ in 
${\mathbb C}_+$ (generic case), 
$M^{(x)}(x,z)$ satisfies the  residue conditions 
\begin{subequations}\label{M-x-res}
\begin{align}
& \underset{z=\nu_j}{\operatorname{Res}} M^{(x)(1)}(x,z) = \frac{e^{2i\nu_j x}b^*(\nu_j)}{\dot a(\nu_j)} 
M^{(x)(2)}(x,\nu_j), \label{M-x-res-1}\\
& \underset{z=\bar\nu_j}{\operatorname{Res}} M^{(x)(2)}(x,z) = 
-\frac{e^{-2i\bar\nu_j x}b(\bar\nu_j)}{\dot a^*(\bar \nu_j)} 
M^{(x)(1)}(x,\bar\nu_j).
\label{M-x-res-2}
\end{align}
\end{subequations}

\item
$M^{(x)}(x,z)\to I$ as $z\to\infty$ for all $x\in(0,L)$.
\end{enumerate}
Then $q_0(x)$ can be obtained by $q_0(x)=2i [M^{(x)}_1]_{12}(x)$,
where $M_1^{(x)}(x)$ is involved in the large-$z$ development of $M^{(x)}(x,z)$:
$
M^{(x)}(x,z) = I + \frac{M_1^{(x)}(x)}{z} + \dots
$.

Now we notice that 
the RH problem in Theorem \ref{RHP} taken at $t=0$ can be mapped to the RH problem associated with $q_0(x)$ as follows:
\begin{equation}\label{map-x-0-1}
M^{(x)}(x,z):=M(x,0,z)\cdot \begin{cases}
\begin{pmatrix}
    a^{-1}(z) & 0 \\ 0 & a(z)
\end{pmatrix}
\begin{pmatrix}
1 & -G_1(z)e^{-2izx} \\ 0 & 1
\end{pmatrix}, & z\in I, \\
\begin{pmatrix}
1 & 0 \\  -G(z)e^{2izx} & 1
\end{pmatrix}, & z\in II, \\
\begin{pmatrix}
1 & G^*(z)e^{-2izx} \\ 0 & 1
\end{pmatrix}, & z\in III, \\
\begin{pmatrix}
    a^*(z) & 0 \\ 0 & (a^*)^{-1}(z)
\end{pmatrix}
\begin{pmatrix}
1 & 0 \\  G_1^*(z)e^{2izx} & 1
\end{pmatrix}, & z\in IV.
\end{cases}
\end{equation}
Indeed:
\begin{enumerate}
	\item 
By straightforward calculations, the jump for $M^{(x)}(x,z)$
 across $\mathbb R$ is as in 
\eqref{M-x-jump}--\eqref{J0-x};
\item  
$M^{(x)}(x,z)$ has no jump across $\Sigma^N$.
\item
All the off-diagonal entries in the r.h.s. of \eqref{map-x-0-1}
go to $0$ exponentially fast as $z\to\infty$, $\Im z\ne 0$ for $x\in(0,L)$
(by the first expressions in \eqref{G-R-1} and \eqref{G-R-2} and since $R(z)\to 0$).
\end{enumerate}

Now consider the mapping of the residue conditions.

\noindent (I) 
    For $z\in I$: 
    $M^{(x)}(x,z) = M(x,0,z)\begin{pmatrix}
        \frac{1}{a(z)} & -\frac{G_1(z)}{a(z)}e^{-2izx} \\ 0 & a(z)
    \end{pmatrix}$.
\begin{enumerate}
        \item 
        At a zero $\nu_j$ of $a(z)$,
\begin{equation}\label{res-1}
M^{(x)(1)}(x,z) = M^{(1)}(x,0,z)\frac{1}{a(z)} = \frac{1}{z-\nu_j}C^{(1)}(x) + \dots.
\end{equation}
On the other hand, from $G_1(z)=-a(z)b(z)+a^2(z)R(z)$ it follows that 
 as $z\to\nu_j$,
\[
G_1(z) = -\dot a(\nu_j)(z-\nu_j)b(\nu_j) + O((z-\nu_j)^2)
\]
and thus
\begin{equation}\label{res-2}
\begin{aligned}
M^{(x)(2)}(x,z) =  & - M^{(1)}(x,0,z) \frac{G_1(z}{a(z)} e^{-2izx} +a(z) M^{(2)}(x,0,z) \\
& = \frac{1}{z-\nu_j}C^{(1)}(x) \dot a(\nu_j)(z-\nu_j)b(\nu_j) e^{-2i\nu_j x}
+ O(z-\nu_j) \\
& = C^{(1)}(x) \dot a(\nu_j)b(\nu_j) e^{-2i\nu_j x} + O(z-\nu_j).
\end{aligned}
\end{equation}
It follows (also using $b(\nu_j)=\frac{1}{b^*(\nu_j)}$) that  
\[
C^{(1)}(x) =e^{2i\nu_j x} \frac{b^*(\nu_j)}{\dot a(\nu_j)}
M^{(x)(2)}(x,\nu_j).
\]
  This, being combined with \eqref{res-1}, gives
  \[
  \underset{z=\nu_j}{\operatorname{Res}}M^{(x)(1)}(x,z) = e^{2i\nu_j x} \frac{b^*(\nu_j)}{\dot a(\nu_j)}
M^{(x)(2)}(x,\nu_j),
  \]
  which is the required  residue condition \eqref{M-x-res-1}.

  \item 

At a pole $\xi_j$ of $R(z)$,
\[
\begin{aligned}
&\underset{z=\xi_j}{\operatorname{Res}}M^{(x)(2)}(x,z) =  -a(\xi_j)\underset{z=\xi_j}{\operatorname{Res}}R(z)e^{-2i\xi_j x}
M^{(1)}(x,0,\xi_j) + a(\xi_j)\underset{z=\xi_j}{\operatorname{Res}} M^{(2)}(x,0,z) \\
& =  -a(\xi_j)\underset{z=\xi_j}{\operatorname{Res}}R(z)e^{-2i\xi_j x}
M^{(1)}(x,0,\xi_j) + a(\xi_j)\underset{z=\xi_j}{\operatorname{Res}}R(z)e^{-2i\xi_j x}
M^{(1)}(x,0,\xi_j) = 0,
\end{aligned}
\]
where we have used \eqref{M-R-res-3}.
    \end{enumerate}


\noindent (II)
For $z\in II$: 
    $M^{(x)}(x,z) = M(x,0,z)\begin{pmatrix}
        1 & 0 \\ -G(z)e^{2izx}  & 1
    \end{pmatrix}$.
In particular, 
\[
M^{(x)(2)}(x,\nu_j) =M^{(2)}(x,0,\nu_j),
\]
where 
$\nu_j$ is a zero of $a(z)$.
 On the other hand, by \eqref{G-R-1},
 \[
\underset{z=\nu_j}{\operatorname{Res}}G(z) = -\frac{b^*(\nu_j)}{\dot a(\nu_j)}
 \]
 and thus
 \[
 \begin{aligned}
\underset{z=\nu_j}{\operatorname{Res}} M^{(x)(1)}(x,z) & = -\underset{z=\nu_j}{\operatorname{Res}}G(z) e^{2i\nu_j x} M^{(2)}(x,0,\nu_j) =-\frac{b^*(\nu_j)}{\dot a(\nu_j)} e^{2i\nu_j x} M^{(x)(2)}(x,\nu_j),
 \end{aligned}
 \]
 which is again the required residue condition \eqref{M-x-res-1}.
 Similarly for \eqref{M-x-res-2}.

Summarizing, transformation \eqref{map-x-0-1} produces
$M^{(x)}(x,z)$ that satisfies the jump and residue conditions for the RH problem associated with $q_0(x)$, which implies that $q(x,0)=q_0(x)$.

\subsubsection{Verifying the periodicity}

In order to verify the periodicity, it is sufficient to relate  the RH problem for $M(L,t,z)$
to that for $M(0,t,z)$ in such a way that both the jump  residue
conditions match correctly. 

Introduce the piece-wise analytic matrix functions: 
\begin{equation}\label{P-t}
    P^{(t)}(t,z)=\begin{cases}
        I, & z\in I\cup IV,\\
        \begin{pmatrix}
            a(z)+b(z)R^*(z) & -b(z)e^{-4iz^2t} \\ 0 & a^*(z)+b^*(z)R(z)
        \end{pmatrix}, & z\in II,\\
        \begin{pmatrix}
            a(z)+b(z)R^*(z) & 0 \\ b^*(z)e^{4iz^2t} & a^*(z)+b^*(z)R(z)
        \end{pmatrix}, & z\in III,
    \end{cases}
\end{equation}
and 
\begin{equation}\label{hat-P-t}
    \hat P^{(t)}(t,z)=\begin{cases}
        I, & z\in II\cup III,\\
        \begin{pmatrix}
            a^*(z)+b^*(z)R(z)  & 0 \\ -b^*(z)e^{2izL}e^{4iz^2t} & a(z)+b(z)R^*(z)
        \end{pmatrix}, & z\in I,\\
        \begin{pmatrix}
            a^*(z)+b^*(z)R(z) &  b(z)e^{-2izL}e^{-4iz^2t} \\ 0 & a(z)+b(z)R^*(z)
        \end{pmatrix}, & z\in IV.
    \end{cases}
\end{equation}
Then introduce 
\begin{equation}\label{M-t}
M^{(t)}(t,z) := M(0,t,z) P^{(t)}(t,z), \qquad 
\hat M^{(t)}(t,z) := M(L,t,z) \hat P^{(t)}(t,z).
\end{equation}

\begin{prop}\label{prop-t}
  $M^{(t)}(t,z)\equiv\hat M^{(t)}(t,z)$; 
  consequently, $q(0,t)=q(L,t)$ and $q_x(0,t)=q_x(L,t)$ for all $t$.
\end{prop}

To prove the proposition, it is sufficient to prove that 
$M^{(t)}(t,z)$ and $\hat M^{(t)}(t,z)$
satisfy the same jump 
  and residue conditions.

1. Using the definitions of $R$, $G$ and $G_2$ as well as the properties
\eqref{R-R-star} and \eqref{a-b-R}
of $R$, it 
 is by straightforward calculations that for $z\in {\mathbb R}\cup i{\mathbb R}$,
 \[
M^{(t)+}(t,z) = M^{(t)-}(t,z)J^{(t)}(t,z) \ \text{   and   }\ 
\hat M^{(t)+}(t,z) = \hat M^{(t)-}(t,z)J^{(t)}(t,z)
 \]
involving the same $J^{(t)}(t,z) = \begin{pmatrix}
     1 & R(z)e^{-4iz^2 t } \\ R^*(z)e^{4iz^2 t } & 1+R^*(z)R(z)
 \end{pmatrix}$, where ${\mathbb R}\cup i{\mathbb R}$ is oriented such that 
 quadrants I and III have positive boundaries.

2. Using, additionally, property \eqref{R-pm-a-b}, it follows that 
on parts of $\Sigma^N$, $J^{(t)}$ is given by 
\[
J^{(t)}(t,z) = \begin{cases}
    \begin{pmatrix}
        1 & (R^+(z)-R^-(z))e^{-4iz^2 t } \\
        0 & 1
    \end{pmatrix}, & z\in I\cup III, \\
    \begin{pmatrix}
        1 &  0 \\
        (R^{*+}(z)-R^{*-}(z))e^{4iz^2 t } 
        & 1
    \end{pmatrix}, & z\in II\cup IV.
\end{cases}
\]

3. In order to prove that 
 $M^{(t)}(t,z)$   and $\hat M^{(t)}(t,z)$ satisfy 
 the same residue conditions, we observe from 
 \eqref{R-equation} that if $\xi_j$ is a pole of $R(z)$, then 
$b^*(\xi_j)=0$
 and 
\begin{equation}\label{poles1}
    \lim_{z\to\xi_j}b^*(z)R(z) = e^{-2izL}a(\xi_j) - a^*(\xi_j);
\end{equation}
consequently,
\begin{equation}\label{poles2}
(a^* + b^*R)|_{z=\xi_j} = e^{-2i\xi_j L}a(\xi_j), \qquad
(a + bR^*)|_{z=\xi_j} = e^{2i\xi_j L}a^*(\xi_j).
\end{equation}
Similarly, if $\xi_j$ is a pole of $R^*(z)$, then 
$b(\xi_j)=0$
 and 
\begin{equation}\label{poles3}
    \lim_{z\to\xi_j}b(z)R^*(z) = e^{2i\xi_j L}a^*(\xi_j) - a(\xi_j)
\end{equation}
whereas \eqref{poles2} keep holding. Using these properties, 
it is again by straightforward calculations that 
$M^{(t)}(t,z)$ and $\hat M^{(t)}(t,z)$ satisfy the same residue conditions:
\begin{enumerate}
    \item 
    At the poles $\xi_j$ of $R(z)$ for $z\in I$ and $z\in III$:
    \begin{equation}\label{M-t-R-res-1}
		\underset{z=\xi_j}{\operatorname{Res}} M^{(t)(2)}(t,z) = e^{-4i\xi_j^2 t}
		\underset{z=\xi_j}{\operatorname{Res}}R(z) M^{(t)(1)}(t,\xi_j).
		\end{equation}
	\item
 At the poles $\xi_j$ of $R^*(z)$ for $z\in II$ and $z\in IV$:
	\begin{equation}\label{M-t-R-res-2}
		\underset{z=\xi_j}{\operatorname{Res}} M^{(t)(1)}(t,z) = -e^{4i\xi_j^2 t}
		\underset{z=\xi_j}{\operatorname{Res}}R^*(z) M^{(t)(2)}(t,\xi_j).
		\end{equation}
  \end{enumerate}

\subsection{From the basic RH problem to a RH problem with structure \eqref{J_j}-\eqref{RHnorm-Psi}
}

The reduction of the basic RH problem to the RH problem with structure \eqref{J_j}-\eqref{RHnorm-Psi}
as in Proposition  \ref{p2} 
can be performed in several consecutive steps.

In Step 1, we ``undress'' the jump matrices on ${\mathbb R}\cup i{\mathbb R}$
to those having a diagonal structure. This step will require 
appropriate \emph{algebraic} factorizations of the jumps.

In Step 2, we reduce the RH problem obtained at Step 1 to that (i) with the contour $\Sigma^N$ only and (ii) having no residue conditions.
This step will require \emph{analytic} factorization of a scalar function.

In Step 3, we reduce the RH problem obtained at Step 2 to that having the structure as in Proposition  \ref{p2}, i.e., involving only constants $\phi_j$ as non-trivial 
elements in the construction of the jump matrices across $\Sigma^N$.

\subsubsection{Step 1: Undressing the jump matrices on ${\mathbb R}\cup i{\mathbb R}$}

Recall that in all our RH problem transformations involving multiplication from the right,
we need that the diagonal part of  the factors approaches the identity matrix 
as $z\to\infty$ (in all domains) whereas the off-diagonal parts decay exponentially 
fast to $0$ for all $t>0$ and all $x\in (0,L)$. Since the off-diagonal parts 
involve $e^{2izx +4iz^2 t}$ or $e^{-2izx -4iz^2 t}$, it follows that the appropriate
factors should have triangular form, with a single non-zero off-diagonal
entry containing the decaying exponential.  

Introduce
\[
D_1(z)=\frac{1}{1+R^*(z)R(z)},\quad D_2(z)=D_1^{-1}(z)=1+R^*(z)R(z),
\]
\[
D_3(z)=\frac{1+R^*(z)R(z)}{a(z)+b(z)R^*(z)}, 
\quad D_4(z)=D_3^*(z)=\frac{1+R^*(z)R(z)}{a^*(z)+b^*(z)R(z)},
\]
\[
U(z)=-\frac{e^{2izL}R(z)}{1+R^*(z)R(z)}, \quad
L(z)=\frac{R^*(z)}{1+R^*(z)R(z)}.
\]

\begin{prop}\label{triang-factor}
    The jump matrix $J_0(z)$ defined by \eqref{J0-2}
    allows the following algebraic factorizations:
\begin{equation}\label{J0-factor}
	J_0(z) = \begin{cases}
	\begin{pmatrix}
	1 & L^*(z) \\ 0 & 1
	\end{pmatrix}
	\begin{pmatrix}
	D_1(z) & 0 \\ 0 & D_1^{-1}(z)
	\end{pmatrix}
	\begin{pmatrix}
	1  & 0 \\ L(z) & 1
	\end{pmatrix}, & z\in {\mathbb R}_+,\\
	\begin{pmatrix}
	1 & 0 \\ U^*(z) & 1
	\end{pmatrix}
	\begin{pmatrix}
	D_2(z) & 0 \\ 0 & D_2^{-1}(z)
	\end{pmatrix}
	\begin{pmatrix}
	1  & U(z) \\ 0 & 1
	\end{pmatrix}, & z\in {\mathbb R}_-,\\
	\begin{pmatrix}
	1  & 0 \\ -L(z) & 1
	\end{pmatrix}
 \begin{pmatrix}
            D_3(z) & 0 \\ 0 &  D_3^{-1}(z)
        \end{pmatrix}
	\begin{pmatrix}
	1  & U(z) \\ 0  & 1
	\end{pmatrix}
 , & z\in i{\mathbb R}_+,\\
	\begin{pmatrix}
	1  & 0 \\ U^*(z) & 1
	\end{pmatrix}
 \begin{pmatrix}
            D_4(z) & 0 \\ 0 &  D_4^{-1}(z)
        \end{pmatrix}
	\begin{pmatrix}
	1  & -L^*(z) \\ 0 & 1
	\end{pmatrix}, & z\in i{\mathbb R}_-.
	\end{cases}
	\end{equation}    
\end{prop}
 \emph{Proof}: by straightforward calculations.

Factorizations \eqref{J0-factor} suggest the undressing transformation
of the RH problem as follows:
\begin{equation}\label{undress}
    \tilde M(x,t,z)= M(x,t,z)\cdot \begin{cases}
        \begin{pmatrix}
	1  & 0 \\ -L(z)e^{2izx+4iz^2t} & 1
	\end{pmatrix}, & z\in I\\
 \begin{pmatrix}
	1  & -U(z)e^{-2izx-4iz^2t} \\ 0 & 1
	\end{pmatrix}, & z\in II\\
 \begin{pmatrix}
	1  & 0 \\ U^*(z)e^{2izx+4iz^2t} & 1
	\end{pmatrix}, & z\in III\\
 \begin{pmatrix}
	1  & L^*(z)e^{-2izx-4iz^2t} \\ 0 & 1
	\end{pmatrix}, & z\in IV.
    \end{cases}
\end{equation}
Notice that this transformation is appropriate in the sense that all the off-diagonal entries in the factors in \eqref{undress} decay exponentially fast to 
$0$ as $z\to\infty$ for all $t>0$ and $x\in (0,l)$.

The jump conditions for $\tilde M$ across ${\mathbb R}\cup i{\mathbb R}$ involve obviously
the diagonal matrices from the r.h.s. of \eqref{J0-factor}. Concerning the jump conditions for $\tilde M$ across $\Sigma^N$ and the residue conditions for the RH problem for $\tilde M$, we have the following two propositions.

\begin{prop}
  $\tilde M$ satisfies the following jump conditions across $\Sigma^N$:
  
    $\tilde M^+(x,t,z) = \tilde M^-(x,t,z)\tilde J(x,t,z)$,
    where 
    $\tilde J(x,t,z) = e^{-(izx + 2iz^2 t)\sigma_3}\tilde J_0(z) e^{(izx + 2iz^2 t)\sigma_3}$
    with 
        \begin{equation}\label{tilde-J0}
	\tilde J_0(z) = \begin{cases}
		\begin{pmatrix}
	0  & R^+(z)-R^-(z) \\ \frac{1}{R^-(z)-R^+(z)} & 0
	\end{pmatrix}, & z\in \Sigma^N\cap I, \\
	\begin{pmatrix}
	0  & \frac{e^{2izL}}{R^{*+}(z)-R^{*-}(z)} \\ (R^{*-}(z)-R^{*+}(z))e^{-2izL} & 0
	\end{pmatrix}, & z\in \Sigma^N\cap II, \\
	\begin{pmatrix}
	0  & (R^+(z)-R^-(z))e^{2izL} \\
 \frac{e^{-2izL}}{R^-(z)-R^+(z)}  & 0
	\end{pmatrix},  & z\in \Sigma^N\cap III, \\
	\begin{pmatrix}
	0  & \frac{1}{R^{*+}(z)-R^{*-}(z)} \\
 R^{*-}(z)-R^{*+}(z) & 0
	\end{pmatrix}, & z\in \Sigma^N\cap IV.
	\end{cases}
	\end{equation}
\end{prop}
\noindent\emph{Proof}. Consider $\Sigma^N\cap I$; here we have
\[
\begin{aligned}
\tilde J_0(z)= & \begin{pmatrix}
    1 & 0 \\ L^- & 1
\end{pmatrix}
\begin{pmatrix}
    1 & R^+ - R^- \\ 0 & 1
\end{pmatrix}
\begin{pmatrix}
    1 & 0 \\ -L^+ & 1
\end{pmatrix} \\
 & = \begin{pmatrix}
    1 & 0 \\ \frac{R^{*-}}{1+R^{*-}R^-} & 1
\end{pmatrix}
\begin{pmatrix}
    1 & R^+ - R^- \\ 0 & 1
\end{pmatrix}
\begin{pmatrix}
    1 & 0 \\ -\frac{R^{*+}}{1+R^{*+}R^+} & 1
\end{pmatrix} \\
& = \begin{pmatrix}
    1 & R^+ - R^- \\ \frac{R^{*-}}{1+R^{*-}R^-} & 1+ \frac{(R^+ - R^-)R^{*-}}{1+R^{*-}R^-}
\end{pmatrix}
\begin{pmatrix}
    1 & 0 \\ -\frac{R^{*+}}{1+R^{*+}R^+} & 1
\end{pmatrix}.
\end{aligned}
\]
Now we notice that the $(22)$ entry in the first matrix equals $0$, because 
\begin{equation}\label{R-pm-2}
    1+R^{*-}R^- + (R^+ - R^-)R^{*-} = 1+R^{*-}R^- + R^+ R^{*-} - R^-R^{*-} = 1+ R^+ R^{*-} = 0
\end{equation}
due to \eqref{R-pm}. It follows that 
\[
\tilde J_0(z) = \begin{pmatrix}
    1 & R^+ - R^- \\ \frac{R^{*-}}{1+R^{*-}R^-} & 0
\end{pmatrix}
\begin{pmatrix}
    1 & 0 \\ -\frac{R^{*+}}{1+R^{*+}R^+} & 1
\end{pmatrix} = \begin{pmatrix}
    0 & R^+ - R^- \\ \frac{1}{R^- - R^+} & 0
\end{pmatrix},
\]
where we have again used the equality \eqref{R-pm-2}.
Similarly for other quadrants.

\begin{prop}\label{tilde-sing}
    $\tilde M$ satisfies the following singularity conditions:
    \begin{enumerate}
        \item 
        For $z\in I\cup III$, 
        \begin{equation}\label{sing-1-3}
            \tilde M(x,t,z)=M_{reg}(x,t,z)\begin{pmatrix}
                z-\xi_j & 0 \\ 0 & \frac{1}{z-\xi_j}
            \end{pmatrix}
        \end{equation}
        at all poles of $R(z)$ in $I\cup III$, where $M_{reg}$ is nowhere singular in $I\cup III$.
        \item 
        For $z\in II\cup IV$, 
        \begin{equation}\label{sing-2-4}
            \tilde M(x,t,z)=M_{reg}(x,t,z)\begin{pmatrix}
              \frac{1}{z-\xi_j}   & 0 \\ 0 & z-\xi_j
            \end{pmatrix}
        \end{equation}
        at all poles $\xi_j$ of $R^*(z)$ in $II\cup IV$, where $M_{reg}$ is nowhere singular
        in $II\cup IV$.
    \end{enumerate}
\end{prop}
\begin{rem}
    The poles of $R^*(z)$ in $II\cup IV$ are complex conjugated to those of $R(z)$ in $I\cup III$.
\end{rem}

\noindent\emph{Proof of Proposition \ref{tilde-sing}}.
Consider $z\in I$, where $\tilde M = M\begin{pmatrix}
    1 & 0 \\ -\frac{R^*}{1+R^*R}e^{2izx+4iz^2t} & 1
\end{pmatrix}$. It follows that $\tilde M^{(2)}=M^{(2)}$ and thus $\tilde M^{(2)}$ has the required 
singularity from \eqref{sing-1-3} due to \eqref{M-R-res-3}.

Now we need to show that $\tilde M^{(1)}(z)=O(z-\xi_j)$ as $z\to\xi_j$. Indeed,  
\[
\begin{aligned}
\tilde M^{(1)}(z) &=   M^{(1)}(z) -\frac{R^*(z)}{1+R^*(z)R(z)}e^{2izx+4iz^2t}M^{(2)}(z) = M^{(1)}(z) \\
 & -\frac{R^*(\xi_j)}{1+R^*(\xi_j)\left(\frac{\underset{z=\xi_j}{\operatorname{Res}}R(z)}{z-\xi_j} + O(1)\right)}\left(\frac{\underset{z=\xi_j}{\operatorname{Res}}R(z)}{z-\xi_j}M^{(1)}(\xi_j) + O(1)\right) = O(z-\xi_j).
\end{aligned}
\]
Similarly for other quadrants.

Looking at the diagonal factors in \eqref{J0-factor}, we notice that we can simplify them
getting rid of $a+bR^*$ and $a^*+b^*R$ by introducing 
\[
\hat M = \tilde M \cdot \begin{cases}
    I, & z\in II\cup III,\\
\begin{pmatrix}
    a^*+b^*R & 0 \\ 0 & a+bR^*
\end{pmatrix}, & z\in I\cup IV.
\end{cases}
\]

Recall that $a+bR^*$ and $a^*+b^*R$ have neither zeros no singularities, and thus $\hat M$ satisfies the same singularity conditions as $\tilde M$.

On the other hand, the jump conditions for $\hat M$ on $\Sigma^N\cap I$
become:
\begin{equation}\label{tilde-Ga-jump}
    \hat J_0 = \begin{pmatrix}
    a+bR^{*-} & 0 \\ 0 & a^*+b^*R^-
\end{pmatrix}\begin{pmatrix}
	0  & R^+(z)-R^-(z) \\ \frac{1}{R^-(z)-R^+(z)} & 0
	\end{pmatrix}\begin{pmatrix}
    a^*+b^*R^+ & 0 \\ 0 & a+bR^{*+}
\end{pmatrix}.
\end{equation}
Using $(a+bR^{*-})(a^*+b^*R^+)=e^{2izL}$ (see \eqref{R-pm-a-b}
and \eqref{a-b-R}), jump \eqref{tilde-Ga-jump}
becomes:
\begin{equation}\label{tilde-Ga-jump-1}
    \hat J_0(z) = \begin{pmatrix}
	0  & (R^+(z)-R^-(z))e^{2izL} \\ \frac{e^{-2izL}}{R^-(z)-R^+(z)} & 0
	\end{pmatrix},\qquad z\in \Sigma^N\cap I,
\end{equation}
which has the same form as for $z\in \Sigma^N\cap III$, see \eqref{tilde-J0}.

Similarly, the jump for $\hat M$ on $\Sigma^N\cap IV$ has the same expression
as that for $\tilde M$ on $\Sigma^N\cap IV$, see \eqref{tilde-J0}.

Changing the orientation of ${\mathbb R}_+$ (setting it to go from $+\infty$ to $0$)
and summarizing, we arrive at the following
\begin{theorem}\label{thm:step-1}
Assuming that the number of the main spectrum points associated with $q_0(x)$ is finite, 
    the solution $q(x,t)$ of the periodic IBVP \eqref{periodic-NLS} 
    can be given by
    $q(x,t)=2i [\hat M_1]_{12}(x,t)$,
where $\hat M_1(x,t)$ enters the large-$z$ development of $\hat M(x,t,z)$:
$
\hat M(x,t,z) = I + \frac{\hat M_1(x,t)}{z} +\dots
$, and $\hat M(x,t,z)$ is  the solution of the following  RH problem: 
given $R(z)$ (which is constructed by \eqref{R-1-2-M} from the scattering coefficients $a(z)$ and $b(z)$
associated with the initial data $q_0(x)$, where the branch is chosen such that 
$R(z)\to 0$ as $z\to\infty$), find $\hat M(x,t,z)$ satisfying the following conditions:
\begin{enumerate}
    \item 
    $\hat M(x,t,z)$ is meromorphic in ${\mathbb C}\setminus ({\mathbb R}\cup i{\mathbb R}\cup\Sigma^N)$;
    \item 
    $\hat M(x,t,z)$ satisfies the jump conditions
    $\hat M^+(x,t,z) = \hat M^-(x,t,z)\hat J(x,t,z)$, where 
    $\hat J(x,t,z) = e^{-(izx + 2iz^2 t)\sigma_3}\hat J_0(z) e^{(izx + 2iz^2 t)\sigma_3}$
    and 
    \begin{equation}\label{hat-jump}
        \hat J_0(z) = \begin{cases}
            \begin{pmatrix}
                1+R^*(z)R(z) & 0 \\ 0 & \frac{1}{ 1+R^*(z)R(z)}
            \end{pmatrix}, & z\in {\mathbb R}\cup i{\mathbb R},\\
            \begin{pmatrix}
	0  & (R^+(z)-R^-(z))e^{2izL} \\ \frac{e^{-2izL}}{R^-(z)-R^+(z)} & 0
	\end{pmatrix}, & z\in \Sigma^N\cap (I\cup III), \\
 \begin{pmatrix}
	0  & \frac{e^{2izL}}{R^{*+}(z)-R^{*-}(z)} \\ (R^{*-}(z)-R^{*+}(z))e^{-2izL} & 0
	\end{pmatrix}, & z\in \Sigma^N\cap (II\cup IV)
        \end{cases}
    \end{equation}
    \item
    $\hat M(x,t,z)$ satisfies the singularity  conditions \eqref{sing-1-3} (for  $z\in I\cup III$) 
    and \eqref{sing-2-4} (for $z\in II\cup IV$).
        \item $\hat M(x,t,z)\to I$ as $z\to \infty$.
\end{enumerate}
\end{theorem}

\subsubsection{$R(z)$ in connection with the theory of periodic finite-band solutions of the NLS}

Before passing to Step 2 (getting rid of jumps across ${\mathbb R}\cup i{\mathbb R}$ 
as well as of the singularity conditions), let us take a look at $R(z)$ taking into account
the connection to the theory of finite-band periodic solutions
(see, e.g., \cite{wahls} and references therein). 

Fix $(x_0,t_0)=(0,0)$ and denote $q=q(0,0)=q_0(0)$.
\begin{prop}\label{g-h-f}
    There exists an entire function $C(z)$ such that
    $\mathcal{M}_{12}(z)=C(z)g(z)$, $\mathcal{M}_{21}(z)=C(z)h(z)$, and 
    $\frac{i}{2}(\mathcal{M}_{11}(z)-\mathcal{M}_{22}(z))= C(z)F(z)$,
        where 
    \begin{enumerate}
                \item
        $g(z)=q \prod\limits_{j=1}^N (z-\mu_j)$ and 
        $h(z)=-\bar q \prod\limits_{j=1}^N (z-\bar\mu_j)$;
        \item
   The points $\{\mu_j\}_1^N$ constitutes the \emph{auxiliary spectrum} (at $(x,t)=(0,0)$) of the Zakharov--Shabat 
   operator \eqref{Lax-x} with a periodic, finite-genus potential $q_0(x)$; it consists of the
   simple zeros of $\mathcal{M}_{12}(z)$ (or $b(z)$), which are not double zeros of  $\Delta^2(z)-4$,
   where $\Delta(z)=\mathcal{M}_{11}(z)+\mathcal{M}_{22}(z)$, and  of the multiple zeros of $\mathcal{M}_{12}(z)$, if any.
   \item 
         $F(z)$ is a polynomial such that the following relation holds:
       \begin{equation}\label{f-w-gh}
           F^2(z)=P(z)+g(z)h(z),
       \end{equation}
       where $P(z)=w^2(z)=\prod\limits_{j=0}^N (z-z_j)(z-\bar z_j)$ and 
         $z_j$ and $\bar z_j$ are simple zeros of $\Delta^2(z)-4$.
       This implies 
       \begin{equation}
           C^2(z)P(z) = \frac{1}{4}(4-\Delta^2(z)).
       \end{equation}
            \end{enumerate}
\end{prop}

In view of Proposition \ref{g-h-f}, $R(z)$ can be expressed as follows:
\begin{itemize}
    \item 
    If $F(z)=z^{N+1}+\dots$, then (recalling that $\sqrt{P(z)}\sim z^{N+1}$ as $z\to\infty$)
    \begin{equation}\label{R1}
    \begin{aligned}
      &R(z) = R_{(1)}(z) = \frac{1}{\mathcal{M}_{21}(z)}\left(
        \frac{\mathcal{M}_{11}(z)-\mathcal{M}_{22}(z)}{2} - \sqrt{\frac{\Delta^2(z)}{4}-1}
        \right) \\
         & = \frac{1}{C(z)h(z)}\left(-iC(z)F(z)+\sqrt{-C^2(z)P(z)}\right)
         = -\frac{i}{h(z)}(F(z)-\sqrt{P(z)}).
    \end{aligned}
\end{equation}
Accordingly,
\begin{equation}\label{R-conj}
    R^{(1)*}(z) = -\frac{i}{g(z)}(F(z)-\sqrt{P(z)})
\end{equation}
(notice that $f^*=f$, $P^*=P$, $C^*=C$, $g^*=-h$)
and 
\begin{equation}\label{R2}
    R_{(2)}(z) = -\frac{1}{R^{(1)*}(z)} = -\frac{i}{h(z)}(F(z)+\sqrt{P(z)}).
\end{equation}
\item 
    If $F(z)=-z^{N+1}+\dots$, then
\begin{equation}\label{RR}
    \begin{aligned}
      & R_{(1)}(z) = -\frac{i}{h(z)}(F(z)+\sqrt{P(z)}).
    \end{aligned}
    \end{equation}
\end{itemize}

\begin{rem}
Concerning the singularity conditions in Theorem \ref{thm:step-1} we observe the following:
\begin{enumerate}
    \item 
    The set of poles of $R(z)$ consists of those zeros of $b^*(z)$ (i)  which  
    are zeros of $h(z)$ (i.e., belong to 
the conjugated auxiliary spectrum $\{\bar\mu_j\}_1^N$),  and, at the same time, (ii) which
are not zeros of $F(z)\mp\sqrt{P(z)}$ (or, in view of \eqref{f-w-gh}, are zeros of 
$F(z)\pm\sqrt{P(z)}$).  Actually, the auxiliary spectrum $\{\bar\mu_j\}_1^N$ consists of 
all poles of $R$ as a function on the two-sheet Riemann surface
    (or, equivalently,  the set of all poles of $R_{(1)}(z)$ and $R_{(2)}(z)$
    as functions on the complex plane).
    \item
    Not all poles of $R(z)$ are involved in the singularity conditions (only those in $I$ and $III$). 
\end{enumerate}
Consequently, in particular cases it is possible that there are no singularity conditions at all
but in general, there can be up to $N$ singularity conditions in $I$ and $III$.
\end{rem}


From \eqref{R1},\eqref{R-conj} and \eqref{RR} it follows that 
\begin{equation}\label{R*R}
  1+R^*(z)R(z) = \frac{2\sqrt{P(z)}}{\tilde F(z)+\sqrt{P(z)}},
\end{equation}
where
\begin{equation}\label{tildeF}
    \tilde F(z)=\begin{cases}
         F(z), & \text{if}\ F(z)=z^{N+1}+\dots, \\
         -F(z), & \text{if}\ F(z)=-z^{N+1}+\dots.
     \end{cases}
 \end{equation}

Thus the zeros (of order $1/2$) of $1+R^*(z)R(z)$ are, generically, the branch points
(the main spectrum points) $\{z_j,\bar z_j\}_0^N$.


\subsubsection{Step 2: getting rid of the jumps across ${\mathbb R}\cup i{\mathbb R}$ as well as of
the singularities}

Due to \eqref{hat-jump}, one can  get rid of jumps across ${\mathbb R}\cup i{\mathbb R}$
by multiplication from the right by  diagonal matrices  $\begin{pmatrix}
    f(z) & 0 \\ 0 & f^{-1}(z)
\end{pmatrix}$, where $f(z)$ is related to a ``square root'' of  $1+R^*(z)R(z)$.

Consider first the particular case, 
assuming  that $R(z)$ has no poles in the whole plane
(particularly, this implies that there are no singularity conditions).
Define
\begin{equation}
    f(z)=\begin{cases}
        (1+R^*(z)R(z))^{\frac{1}{2}}, & z\in (I\cup III)\setminus \Sigma^N, \\
        (1+R^*(z)R(z))^{-\frac{1}{2}}, & z\in (II\cup IV) \setminus \Sigma^N ,
    \end{cases}
\end{equation}
such that $f(z)\to 1$ as $z\to\infty$,
and introduce
\begin{equation}
    \check M(x,t,z)= \hat M(x,t,z) \begin{pmatrix}
    f(z) & 0 \\ 0 & f^{-1}(z)
\end{pmatrix}, \qquad z\in {\mathbb C}\setminus ({\mathbb R}\cup i {\mathbb R} \cup \Sigma^N).
\end{equation}
Using \eqref{hat-jump},  direct calculations give that $\check M(x,t,z)$ has no jumps across ${\mathbb R}\cup i {\mathbb R}$.

Now we calculate  $\check J_0(z)$ 
in the jump conditions for $\check M(x,t,z)$ across $\Sigma^N$:
$\check M^+(x,t,z) = \check M^-(x,t,z)\check J(x,t,z)$
with $\check J(x,t,z) = e^{-(izx + 2iz^2 t)\sigma_3}\check J_0(z) e^{(izx + 2iz^2 t)\sigma_3}$. We have
\[
[\check J_0]_{12}(z) = [\hat J_0]_{12}(z)\frac{1}{f_+(z)f_-(z)}.
\]
Consequently, for $z\in \Sigma^N\cap (I\cup III)$ we have
\begin{equation}\label{J-12-1}
    \begin{aligned}
{[\check J_0]}_{12}(z) & = 
    \frac{e^{2izL}(R^+(z)-R^-(z))}{(1+R^{*+}(z)R^+(z))^{\frac{1}{2}}
    (1+R^{*-}(z)R^-(z))^{\frac{1}{2}}} 
     = -e^{2izL}R^-(z)\frac{(1+R^{*+}(z)R^+(z))^{\frac{1}{2}}}{(1+R^{*-}(z)R^-(z))^{\frac{1}{2}}},
    \end{aligned}
\end{equation}
where we have used the equality (following from \eqref{R-pm})
\[
R^- - R^+ = R^-(1+R^{*+}R^+).
\]

Similarly, for $z\in \Sigma^N\cap (II\cup IV)$ we have
\begin{equation}\label{J-12-2}
    \begin{aligned}
{[\check J_0]}_{12}(z) & = \frac{e^{2izL}}{R^{*+}(z)-R^{*-}(z)}
   (1+R^{*+}(z)R^+(z))^{\frac{1}{2}}
    (1+R^{*-}(z)R^-(z))^{\frac{1}{2}} \\
    & = \frac{e^{2izL}}{R^{*+}(z)}\frac{(1+R^{*+}(z)R^+(z))^{\frac{1}{2}}}{(1+R^{*-}(z)R^-(z))^{\frac{1}{2}}} 
    = -e^{2izL}R^-(z)\frac{(1+R^{*+}(z)R^+(z))^{\frac{1}{2}}}{(1+R^{*-}(z)R^-(z))^{\frac{1}{2}}}.
    \end{aligned}
\end{equation}

Thus ${[\check J_0]}_{12}(z) $ has the same analytic expression \eqref{J-12-1}
across all parts of $\Sigma^N$. 
Accordingly,
\begin{equation}\label{J-21-1}
    \begin{aligned}
{[\check J_0]}_{21}(z) = -\left({[\check J_0]}_{12}(z) \right)^{-1} = 
-e^{2izL}R^{*+}(z)\frac{(1+R^{*-}(z)R^-(z))^{\frac{1}{2}}}{(1+R^{*+}(z)R^+(z))^{\frac{1}{2}}}.
\end{aligned}
\end{equation}

\begin{theorem}\label{thm:step-2}
Assuming that $R(z)$ associated with the initial data of a periodic finite-band solution $q(x,t)$
of the NLS equation has no poles, $q(x,t)$ can be given in terms of the solution $\check M(x,t,z)$
of a  RH problem with the jump conditions across $\Sigma^N$ only:
 given $R(z)$, find $\check M(x,t,z)$ satisfying the following conditions:
\begin{enumerate}
    \item 
    $\check M(x,t,z)$ is analytic  in ${\mathbb C}\setminus \Gamma$;
    \item 
    $\check M(x,t,z)$ satisfies the jump conditions across $\Gamma$:
    $\check  M^+(x,t,z) = \check  M^-(x,t,z)\check  J(x,t,z)$, where 
    $\check  J(x,t,z) = e^{-(izx + 2iz^2 t)\sigma_3}\check  J_0(z) e^{(izx + 2iz^2 t)\sigma_3}$
    and 
    \begin{equation}\label{check-jump}
        \check  J_0(z) = 
 \begin{pmatrix}
	0  & iJ_{00}(z) \\ iJ_{00}^{-1}(z) & 0
	\end{pmatrix}, \qquad  z\in \Gamma
    \end{equation}
    with 
    \begin{equation}\label{P}
   J_{00}(z)=ie^{2izL}R^-(z)\frac{(1+R^{*+}(z)R^+(z))^{\frac{1}{2}}}{(1+R^{*-}(z)R^-(z))^{\frac{1}{2}}};     
    \end{equation} 
    
\item $\check M(x,t,z)\to I$ as $z\to \infty$.
\end{enumerate}
Namely, 
    $q(x,t)=2i [\check M_1]_{12}(x,t)$,
where $\check M_1(x,t)$ enters the large-$z$ development of $\check M(x,t,z)$:
$
\check M(x,t,z) = I + \frac{\check M_1(x,t)}{z} +\dots
$. 
\end{theorem}


\begin{rem}\label{rem:4-12}

$J_{00}(z)$ in \eqref{P} looks  complicated,
    but it turns out that its square has a simple expression in terms of $\mathcal{M}_{12}$ and $\mathcal{M}_{21}$
    (or $b(z)$ and $b^*(z)$). Indeed,
    \begin{equation}\label{J-sqr}
        \begin{aligned}
    J_{00}^2(z) & = -e^{4izL}(R^-(z))^2 
    \frac{1+R^{*+}(z)R^+(z)}{1+R^{*-}(z)R^-(z)} = e^{4izL}(R^-(z))^2 \frac{R^+(z)}{R^-(z)} \\
    & = e^{4izL}R^+(z)R^-(z) = -e^{4izL}\frac{R^+(z)}{R^{*+}(z)} = 
    -e^{4izL}\frac{\mathcal{M}_{12}(z)}{\mathcal{M}_{21}(z)}
    = e^{2izL}\frac{b(z)}{b^*(z)},
    \end{aligned}
    \end{equation}
    where we have again used \eqref{R-pm} as well as \eqref{R-R-star}.

\end{rem}

Now consider the general case, where $R(z)$ can have poles. 
Denote by $P_1=\{\xi_j\}$ the set of poles of  $R(z)$ in ${\mathbb C}_+$
and by $P_2=\{\mu_j\}$ the set of poles of  $R^*(z)$ in ${\mathbb C}_+$ (thus the set of all poles of $R(z)$ in the whole ${\mathbb C}$ 
is given by $P_1\cup \bar P_2$, where 
$\bar P_2=\{\bar\mu_j\}$).
Introduce the function $\nu(z)$ having neither zeros no singularities in 
${\mathbb C}_+ \setminus \Sigma^N$:
\begin{equation}\label{nu}
    \nu(z):=\frac{1}{1+R^{*}(z)R(z)} \prod\limits_{j\in P_1}\frac{z-\bar\xi_j}{z-\xi_j}
    \prod\limits_{j\in P_2}\frac{z-\bar\mu_j}{z-\mu_j}.
\end{equation}
Then we can define $\nu^{1/2}(z)$ as an analytic function for $z\in {\mathbb C}_+ \setminus \Sigma^N$
such that $\nu^{1/2}(z)\to 1$ as $z\to\infty$. 

Using $\nu^{1/2}(z)$ we define $f(z)$ in ${\mathbb C} \setminus \Sigma^N$ as follows:
\begin{equation}\label{f-P}
    f(z)=\begin{cases}
        f_I(z)=\nu^{1/2}(z) \prod\limits_{j\in P_2}\frac{z-\mu_j}{z-\bar\mu_j}(1+R^{*}(z)R(z)), & z\in I \\
        f_{II}(z) = \nu^{1/2}(z) \prod\limits_{j\in P_2}\frac{z-\mu_j}{z-\bar\mu_j}, & z\in II\\
        f_{III}(z) = \frac{1}{f^*_{II}(z)}, & z\in III\\
        f_{IV}(z) = \frac{1}{f^*_{I}(z)}, & z\in IV
    \end{cases}
\end{equation}
Then it is straightforward to check that $\check M=\hat M \begin{pmatrix}
    f & 0 \\ 0 & f^{-1}
\end{pmatrix}$
has no singularities in ${\mathbb C} \setminus \Sigma^N$. 
Particularly, for $z\in II$, it is the first column of $\hat M$ that is singular 
at $\mu_j\in P_2$. Then, by the definition of $f$ in $II$, this singularity 
is cancelled for $\check M$. 

For $z\in I$, the second column of $\hat M$  is singular 
at $\xi_j\in P_1$; this singularity is cancelled for $\check M$
since the second column of $\hat M$ is multiplies by $f_I^{-1}(z)$,
which vanishes at such $\xi_j$ due to the factor $(1+R^*(z)R(z))^{-1}$.

By symmetry, the singularities of $\hat M$ in $II\cup IV$ are cancelled 
for $\check M$ as well.

Let us calculate the jump for $\check M$ on $\Sigma^N$.
For $z\in \Sigma^N\cap I$ we have
\begin{equation}\label{J-P-12-1}
    \begin{aligned}
{[\check J_0]}_{12}(z) & = e^{2izL}
    \frac{R^+(z)-R^-(z)}{f_I^+(z) f_I^-(z)} \\
    & = -e^{2izL}R^-(z)\frac{1+R^{*+}(z)R^+(z)}{\nu^{\frac{1}{2}+}(z)\nu^{\frac{1}{2}-}(z)
    \prod\limits_{j\in P_2}\left(\frac{z-\mu_j}{z-\bar\mu_j}\right)^2
    (1+R^{*+}(z)R^+(z))(1+R^{*-}(z)R^-(z))}\\
    & = -e^{2izL}R^-(z)\frac{\nu^{-\frac{1}{2}+}(z)\nu^{-\frac{1}{2}-}(z)}{1+R^{*-}(z)R^-(z)}
     \prod\limits_{j\in P_2}\left(\frac{z-\bar\mu_j}{z-\mu_j}\right)^2.
    \end{aligned}
\end{equation}
Defining $(1+R^{*+}R^+)^{\frac{1}{2}}(1+R^{*-}R^-)^{\frac{1}{2}}$
and $(1+R^{*+}R^+)^{\frac{1}{2}}(1+R^{*-}R^-)^{-\frac{1}{2}}$
in accordance with \eqref{nu}:
\[
(1+R^{*+}(z)R^+(z))^{\frac{1}{2}}(1+R^{*-}(z)R^-(z))^{\frac{1}{2}} := \nu^{-\frac{1}{2}+}(z)\nu^{-\frac{1}{2}-}(z)
\prod\limits_{j\in P_1}\frac{z-\bar\xi_j}{z-\xi_j}
    \prod\limits_{j\in P_2}\frac{z-\bar\mu_j}{z-\mu_j},
\]
\begin{equation}\label{sqrt-1}
    \frac{(1+R^{*+}(z)R^+(z))^{\frac{1}{2}}}{(1+R^{*-}(z)R^-(z))^{\frac{1}{2}}} := 
\frac{\nu^{-\frac{1}{2}+}(z)\nu^{-\frac{1}{2}-}(z)}{1+R^{*-}(z)R^-(z)}
\prod\limits_{j\in P_1}\frac{z-\bar\xi_j}{z-\xi_j}
    \prod\limits_{j\in P_2}\frac{z-\bar\mu_j}{z-\mu_j},
\end{equation}
the expression for ${[\check J_0]}_{12}(z)$ can be written as 
\begin{equation}\label{J-P-12-2}
{[\check J_0]}_{12}(z) = -e^{2izL}R^-(z)\frac{(1+R^{*+}(z)R^+(z))^{\frac{1}{2}}}{(1+R^{*-}(z)R^-(z))^{\frac{1}{2}}}
\prod\limits_{j\in P_1}\frac{z-\xi_j}{z-\bar\xi_j}
    \prod\limits_{j\in P_2}\frac{z-\bar\mu_j}{z-\mu_j}.
\end{equation}

Now we notice that the set $\{\xi_j\}_{j\in P_1}\cup \{\bar\mu_j\}_{j\in P_2}$ is the set of  \emph{all} poles of 
$R(z)$ (in the whole complex plane).
Similarly for $z\in \Sigma^N\cap II$.

For $z\in \Sigma^N\cap III$ we have 
\begin{equation}\label{J-P-12-3}
    \begin{aligned}
{[\check J_0]}_{12}(z) & = e^{2izL}R^-(z) 
    \frac{R^+(z)-R^-(z)}{f_{III}^+(z) f_{III}^-(z)} 
    = - e^{2izL} R^-(z) (1+R^{*+}(z)R^+(z)) f_{II}^{*+}(z) f_{II}^{*-}(z)
    \\
    & = -e^{2izL}R^-(z)(1+R^{*+}(z)R^+(z))\nu^{\frac{1}{2}*+}(z)\nu^{\frac{1}{2}*-}(z)
    \prod\limits_{j\in P_2}\left(\frac{z-\bar\mu_j}{z-\mu_j}\right)^2\\
         & = -e^{2izL}R^-(z)\frac{(1+R^{*+}(z)R^+(z))^{\frac{1}{2}}}{(1+R^{*-}(z)R^-(z))^{\frac{1}{2}}}
\prod\limits_{j\in P_1}\frac{z-\xi_j}{z-\bar\xi_j}
    \prod\limits_{j\in P_2}\frac{z-\bar\mu_j}{z-\mu_j},
    \end{aligned}
\end{equation}
where $\frac{(1+R^{*+}(z)R^+(z))^{\frac{1}{2}}}{(1+R^{*-}(z)R^-(z))^{\frac{1}{2}}}$
is understood as
\begin{equation}\label{sqrt-2}
\frac{(1+R^{*+}(z)R^+(z))^{\frac{1}{2}}}{(1+R^{*-}(z)R^-(z))^{\frac{1}{2}}} := 
\nu^{\frac{1}{2}*+}(z)\nu^{\frac{1}{2}*-}(z)(1+R^{*+}(z)R^-(z))
\prod\limits_{j\in P_1}\frac{z-\bar\xi_j}{z-\xi_j}
    \prod\limits_{j\in P_2}\frac{z-\bar\mu_j}{z-\mu_j},
\end{equation}

Summarizing, on all parts of $\Sigma^N$, the $(12)$ entry of the jump matrix 
$\check M$ has the same analytic expression \eqref{J-P-12-2}, where the square roots 
are understood as \eqref{sqrt-1} in ${\mathbb C}_+$ and as \eqref{sqrt-2} in ${\mathbb C}_-$.

\begin{theorem}\label{thm:step-2-poles}
Let $R(z)$ be associated with the initial data of a periodic finite-band solution $q(x,t)$
of the NLS equation. Denote by $P=\{\xi_j\}$ the set of \emph{all}
poles of $R(z)$ in $\mathbb C$. 
Then $q(x,t)$ can be given in terms of the solution $\check M(x,t,z)$
of the following RH problem: given $R(z)$, find $\check M(x,t,z)$ satisfying the following conditions:
\begin{enumerate}
    \item 
    $\check M(x,t,z)$ is analytic  in ${\mathbb C}\setminus \Sigma^N$;
    \item 
    $\check M(x,t,z)$ satisfies the jump conditions across $\Sigma^N$:
    $\check  M^+(x,t,z) = \check  M^-(x,t,z)\check  J(x,t,z)$, where 
    $\check  J(x,t,z) = e^{-(izx + 2iz^2 t)\sigma_3}\check  J_0(z) e^{(izx + 2iz^2 t)\sigma_3}$
    and 
    \begin{equation}\label{check-jump-poles}
        \check  J_0(z) = 
 \begin{pmatrix}
	0  & iJ_{00}(z) \\ iJ_{00}^{-1}(z) & 0
	\end{pmatrix}, \qquad  z\in \Sigma^N,
    \end{equation}
    with 
    \begin{equation}\label{P-poles}
   J_{00}(z)=ie^{2izL}R^-(z)\frac{(1+R^{*+}(z)R^+(z))^{\frac{1}{2}}}{(1+R^{*-}(z)R^-(z))^{\frac{1}{2}}} \prod\limits_{j\in P}\frac{z-\xi_j}{z-\bar\xi_j}; 
    \end{equation} 
    
\item $\check M(x,t,z)\to I$ as $z\to \infty$.
\end{enumerate}
Namely, 
    $q(x,t)=2i [\check M_1]_{12}(x,t)$,
where $\check M_1(x,t)$ enters the large-$z$ development of $\check M(x,t,z)$:
$
\check M(x,t,z) = I + \frac{\check M_1(x,t)}{z} +\dots
$. 
\end{theorem}

\begin{rem}
    In accordance with Remark \ref{rem:4-12}, 
    \begin{equation}\label{P2-poles}
      J_{00}^2(z)=e^{2izL}\frac{b(z)}{b^*(z)}
      \prod\limits_{j\in P}\left(\frac{z-\xi_j}{z-\bar\xi_j}\right)^2, \qquad  z\in \Sigma^N,
    \end{equation} 
    i.e., up to the sign, the entries of the jump matrix \eqref{check-jump}
   have simple expressions in terms of $b(z)$ and the poles of $R(z)$. 
\end{rem}

\subsubsection{Step 3: Reducing  the jump across $\Sigma^N $ to the form of \eqref{J_j}}

This step can also be performed by multiplication from the right
by an appropriate diagonal matrix. Namely, consider the following 
scalar RH-type problem: given $J_{00}(z)$ for $z\in\Sigma^N$,
find $d(z)$ such that:
\begin{enumerate}
    \item 
    $d(z)$ is analytic in ${\mathbb C}\setminus \Sigma^N$;
    \item
    \begin{equation}\label{d-RHP}
        d_+(z)d_-(z) = J_{00}(z)e^{i\phi_j},\qquad z\in \Sigma_j,\ j=0,\dots,N,
    \end{equation}
    where the constants $\{\phi_j\}_0^N$ are not specified apriori;
    \item
    $d(z)\to 1$ as $z\to\infty$.
\end{enumerate}

Applying the logarithm and dividing by $w(z)=\left(\prod\limits_{j=0}^N (z-z_j)(z-\bar z_j)\right)^\frac{1}{2}$, the problem reduces to the standard additive RH problem,
which gives $d(z)$ satisfying \eqref{d-RHP} in terms of the Cauchy integral:
\begin{equation}\label{d-scalar}
    d(z)=\exp\left\{\frac{w(z)}{2\pi i }\sum\limits_{j=0}^N
    \int_{\Sigma_j}\frac{\log J_{00}(s) +i\phi_j}{w_+(s)(s-z)}ds\right\}.
\end{equation}

Then $\{\phi_j\}$ are determined by applying condition (iii). Indeed, since $w(z)=z^{N+1}+\dots$, by
writing $\frac{1}{s-z}$ as 
\[
\frac{1}{s-z} = -\frac{1}{z}\left(1+\frac{s}{z}+\left(\frac{s}{z}\right)^2 + \dots\right),
\]
we arrive at the requirements that 
\[
\sum\limits_{j=0}^N
    \int_{\Sigma_j}\frac{(\log J_{00}(s) +i\phi_j)s^{l-1}}{w_+(s)}ds = 0,
    \quad l=1,\dots, N+1,
\]
which gives the system of $N+1$ linear equations for $\{\phi_j\}_0^N$:
\begin{equation}\label{phi-eq}
    K \phi = B, \qquad \phi:=(\phi_0, \dots, \phi_N)^T,
\end{equation}
where 
\begin{equation}\label{K-B}
    K_{lm} = \int_{\Sigma_{m-1}}\frac{s^{l-1}}{w_+(s)}ds, \quad
    B_l = i\sum\limits_{j=0}^N \int_{\Sigma_j}
    \frac{\log J_{00}(s)s^{l-1}}{w_+(s)}ds,
    \quad l, \, m=1, \, \dots, \, N+1.
\end{equation}

Then, introducing 
\[
N(x,t,z):=\check M(x,t,z)\begin{pmatrix}
    d(z) & 0 \\ 0 & d^{-1}(z)
\end{pmatrix},
\]
the jump condition reduces to 
 $N^+(x,t,z) = N^-(x,t,z)J_N(x,t,z)$, where 
 \[
J_N(x,t,z) = \begin{pmatrix}
    0 & ie^{-i\phi_j-2izx-4iz^2t} \\ 
    ie^{i\phi_j+2izx+4iz^2t} & 0
\end{pmatrix},\qquad z\in \Sigma_j,
 \]
 i.e., to the form of \eqref{J_j}.

 Thus, we arrived at the following algorithm for solving the direct problem.

\begin{theorem}\label{dir-main}
Let $q(x,t)$ be the finite-band, periodic solution
of the NLS equation, with the spacial period $L$, determined by the real constants
$\{\phi_j\}_0^N$, $\phi_j\in [0,2\pi)$ and constructed by \eqref{q-RH} via the 
solution of the RH problem \eqref{J_j}--\eqref{RHnorm-Psi}. 
Then the constants $\{\phi_j\}_0^N$ can be retrieved from $q(x,0)$, $x\in (0,L)$
via the solution of the system of linear algebraic equations \eqref{phi-eq},
where the coefficients $K$ and $B$ are determined by $J_{00}(z)$ through 
\eqref{K-B}. Here  $J_{00}(z)$ in turn is determined by \eqref{P-poles}
in terms of the spectral function $R(z)$ constructed from $a(z)$ and $b(z)$
associated with $q(x,0)$ as entries of the scattering matrix for the Zakharov--Shabat
equation on the  line with a finitely supported potential $q(x,0)$ (continued by $0$ on the whole axis).
\end{theorem} 
    
\begin{rem}
    From \eqref{d-RHP} we see that the replacement of $J_{00}(z)$ by $-J_{00}(z)$
    on a particular $\Sigma_j$ can be compensated
    by the shift of $\phi_j$ by $\pi$.
    It follows that if we define $J_{00}(z)$
    at each $\Sigma_j$ as any (continuous) branch of the  square root of 
    $J_{00}^2(z)$, then we  can retrieve $\phi_j$ up
    to a shift by $\pi$.
\end{rem}

\section{Evolution}\label{sec:evolution}

In the previous section we have shown that given $q(x,0)$, where $q(x,t)$
is a periodic finite-band solution of the NLS equation, one can
retrieve the underlying ``phases'' $\{\phi_j\}_0^N$ 
(generating $q(x,t)$ through the solution of the RH problem \eqref{J_j}--\eqref{RHnorm-Psi}).

We first notice that the idea of the backward propagation in the spectral terms
using the evolution of the scattering coefficients of the problem on the line:
\[
a(z;T)=a(z;0), \qquad b(z;T)= b(z;0)e^{-4iz^2 t}
\]
does not work in our case since $a(z;T)$ and $b(z;T)$
come from the Jost solutions that are normalized differently compared with
those used for determining $a(z;0)$ and $b(z;0)$.

On the other hand, it is the representation of $q(x,t)$ in terms
of the RH problem \eqref{hatphi-jump}--\eqref{hatphi-norm} that makes it possible
to obtain $\{\phi_j\}_0^N$ from those phases obtained from 
$q(x,T)$ following the procedure presented in Theorem \ref{dir-main}
where $q(x,T)$ is considered as the initial data.

Indeed, let us introduce $\tilde t = t-T$ and let 
$\{\phi_j^T\}_0^N$ be the ``phases'' obtained from $q(x,T)$.
Then, according to \eqref{hatphi-jump}--\eqref{hatphi-norm},
$q(x,t)$ can be obtained 
as 
\begin{equation}\label{q-tilde}
q(x,t) = 2i [\hat\Phi_1^T]_{12}(x,t) e^{2i f_0 x + 2i g_0 \tilde t}
= 2i [\hat\Phi_1^T]_{12}(x,t) e^{2i f_0 x + 2i g_0 t} e^{-2ig_0 T},
\end{equation}
from the solution $\hat \Phi^T$ of the RH problem of type 
\eqref{hatphi-jump}--\eqref{hatphi-norm} with the jump matrices 
\begin{equation}\label{hatj-tilde}
    \hat J_j^T(x,t) = \begin{pmatrix}
    0 & ie^{-i(\phi_j^T+C^f_j x + C^g_j (t-T))} \\
    ie^{i(\phi_j^T+C^f_j x + C^g_j (t-T))} & 0
    \end{pmatrix}.
\end{equation}

Now observe that (i) the expression \eqref{q-tilde} being 
compared with \eqref{q-Rhmod}
contains the factor $e^{-2ig_0 T}$ and (ii)
the multiplication of $q$ by $e^{iC}$ with some real $C$ corresponds to 
the transformation  
$\hat \Phi\mapsto e^{iC/2 \sigma_3}\hat \Phi e^{-iC/2 \sigma_3}$,
which in turn corresponds to the transformation of the jump matrix
$\hat J\mapsto e^{iC/2 \sigma_3}\hat J e^{-iC/2 \sigma_3}$, or, in terms 
of $\hat J_{12}$, to the transformation $\hat J_{12}\mapsto 
\hat J_{12} e^{iC}$. It follows that 
$q(x,t)$ can be expressed exactly as in \eqref{q-Rhmod} in terms of the 
solution of the RH problem with the jump matrix
\[
\begin{pmatrix}
    0 & ie^{-i(\phi_j^T+C^f_j x + C^g_j (t-T)-2g_0 T)} \\
    ie^{i(\phi_j^T+C^f_j x + C^g_j (t-T)-2g_0 T)} & 0
    \end{pmatrix}.
\]
Comparing this with \eqref{hatj} we see that the jumps are the same provided
$\phi_j^T$ and $\phi_j$ are related by
\begin{equation}\label{phi-evol}
    \phi_j = \phi_j^T -(C^g_j + 2g_0)T.
\end{equation}

Expression \eqref{phi-evol} presents the linear evolution of the phases
allowing retrieving the original phases $\phi_j$ (corresponding to 
$t=0$) using the 
phases $\phi_j^T$ obtained as the solution of the direct problem
following the algorithm presented in Theorem \ref{dir-main}
and applied to $q(x,T)$, $x\in(0,L)$ instead of $q(x,0)$.

\section{Examples}\label{sec:examples}
\subsection{Case of $N=1$}

Let us consider a few examples of genus-$1$ case sharing the same 
$z_0$ and $z_1$ but having different phases. For our approach to work, we
need the underlying $q(x,t)$ to be periodic in $x$. According to 
\eqref{q-Rhmod}, in the case $N=1$ 
we have to provide the commensurability
of $f_0$ from \eqref{f} and $C_1^f$ that enters the jump
matrix \eqref{hatj}. A possible way to achieve this is to provide $f_0=0$
by choosing $z_0$ and $z_1$ appropriately.
From \eqref{CfCg} and \eqref{f} it follows that given $z_0$ and $z_1$,
$f_0$ is calculated by
\[
f_0=-\frac{1}{2}\sum_{j=0}^1 (z_j+\bar z_j) + \int_{\Sigma_1}\frac{\xi d\xi}{w_+(\xi)}\left(\int_{\Sigma_1}\frac{d\xi}{w_+(\xi)}\right)^{-1}.
\]
Consequently, starting from some $z_0$ and $z_1$ and calculating
the respective $f_0$, applying the shift $z_j\mapsto z_j+f_0$, $j=0,1$
 produces the needed values of $z_0$ and $z_1$
(generating $f(z)$ with $f_0=0$).

In the following examples, we fix $z_0$ and $z_1$ by
$z_0=0.2780 + i$, $z_1=1.2780 + i$ (for which we have $f_0$ 
to be approximately equal to $0$), take three pairs of 
$\phi_0$ and $\phi_1$, generate $q(x,0)$ by solving the RH problems
\eqref{J_j}--\eqref{RHnorm-Psi}
(we implement the RH problem solver \cite{trogdon2015riemann, olver2012general, olverGit}), and recover $\phi_0$ and $\phi_1$
from $q(x,0)$ 
following the algorithm presented in Theorem
\ref{dir-main}.

According to this algorithm, we have to evaluate $R(z)$ 
from the scattering matrix (or the monodromy matrix) associated
with $q(x,0)$. In this respect, we note that 
in the case $N=1$, an efficient alternative way to 
evaluate $R(z)$ is to use its representation $R(z)=-\frac{i}{h(z)}(F(z)-\sqrt{P(z)})$, see \eqref{R1}, where the coefficients of the polynomials
$F(z)=z^2+a_1 z + a_0$, $g(z)=q(z-\mu)$ and $h(z)=-\overline{q}(z-\overline{\mu})$ (here $q=q(0,0)$) are characterized through \eqref{f-w-gh}:

\begin{align}\label{gen_1_coeff}
    a_1 &= -(\Re z_0 + \Re z_1);\\
    a_0 &= \frac{1}{2}(-(\Re z_0 + \Re z_1)^2+|z_0|^2+|z_1|^2+4 \Re z_0 \Re z_1 -|q|^2);\\
    \Re \mu &= \frac{a_1 a_0+\Re z_1|z_0|^2+\Re z_0|z_1|^2}{|q|^2};\\
    |\Im \mu| &=\sqrt{ \left|\frac{|z_0|^2|z_1|^2-a_0^2}{|q|^2}-(\Re \mu)^2\right|}.
\end{align}
Further, $\Im \mu$ can be specified requiring that $\mathcal{M}_{12}(\mu)=0$.

Then we check whether $\overline{\mu}$ is the pole of $R(z)$
(it is not if $F(\overline{\mu})-\sqrt{P(\overline{\mu})}=0$)
and proceed to constructing $J_{00}(z)$ by \eqref{P} in the case
$R(z)$ has no poles, or by \eqref{P-poles} in the case  when $R(z)$ has a pole. At this point, it is interesting to compare
$J_{00}(z)$ with that obtained as  the principal branch 
of $\sqrt{J_{00}^2(z)}$, where $J_{00}^2(z)$ is given
by simpler formulas, \eqref{J-sqr} or \eqref{P2-poles}, i.e. directly in terms 
of the entries of the scattering matrix.

\textbf{Example 1.} Let  $\phi_0=0.4$ and $\phi_1=0.8$.  
Solving RH problem \eqref{J_j}--\eqref{RHnorm-Psi} gives
$q$, whereas equations \eqref{gen_1_coeff}
give  $a_0$, $a_1$; $\mu$ as shown in Table \ref{table_} ($\Im\mu$ 
 is chosen such that $\mathcal{M}_{12}(\mu)=0$). Thus, the candidate for a pole of $R(z)$ is $z=\bar\mu=0.7780 - 0.3163 i$,
but the direct check shows that $F(\overline{\mu})-\sqrt{P(\overline{\mu})}=0$ and thus $R(z)$ has no poles.
Consequently, in his case $J_{00}(z)$ is given by \eqref{P},
and the direct check shows that it coincides with that 
determined by \eqref{J-sqr} on both bands, $\Sigma_0$ and $\Sigma_1$.

\textbf{Example 2.} Let $\phi_0=0.4+\pi\approx 3.5416$ and $\phi_1=0.8+\pi
\approx 3.9416$.  
Analytically, $q(x,t)$ in this case is that as in Example 1
multiplied by $-1$; the same is for $R(z)$.
As for comparing $J_{00}(z)$ obtained from \eqref{P} and \eqref{J-sqr},
in this case  they are also related by multiplication by $-1$.

\textbf{Example 3.} Let  $\phi_0=0.4$ and $\phi_1=0.8+\pi\approx 3.9416$.  
As above,  $\overline{\mu}$ is not a pole of $R(z)$.
In this case, $J_{00}(z)$ obtained from \eqref{P} and \eqref{J-sqr}
coincide on $\Sigma_0$ and differ by sign on $\Sigma_1$.

\begin{figure}[!h]
    \centering
    \includegraphics[scale=.53]{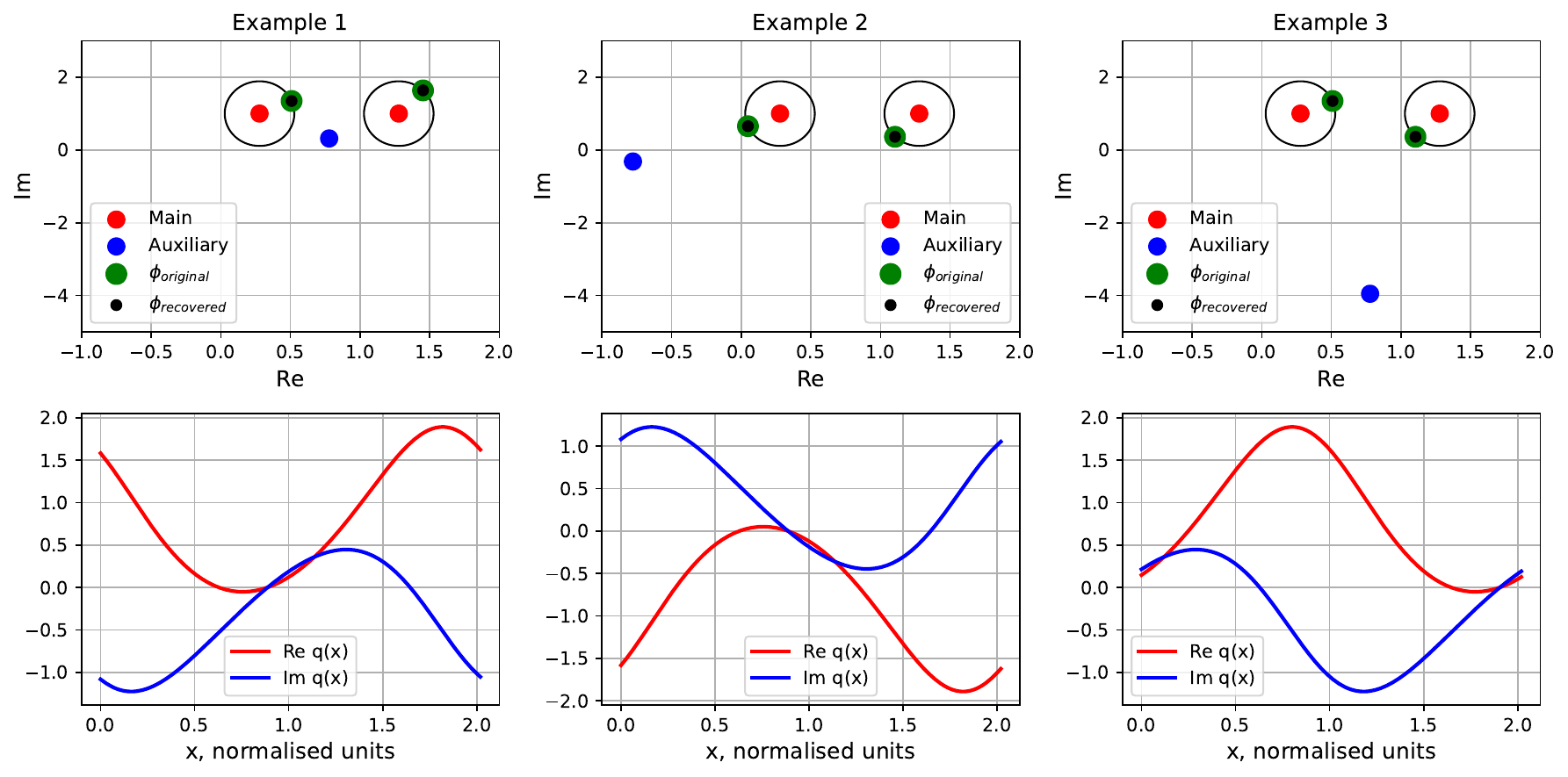}
    \caption{Three examples in the case with $N=1$ with common main spectrum $z_0=0.2780 + i$ and $z_1= 1.2780 + i$ and different phases $\phi_0$ and $\phi_1$. The phases are depicted as points $e^{\phi_j}$ on the unit circles around the corresponding $z_j$.}
    \label{fig:1genus}
\end{figure}

In all three examples, the results of the reconstruction of the phases
are in good agreement with the original $\phi_0$ and $\phi_1$; see also Fig. \ref{fig:1genus}.

\renewcommand{\tabcolsep}{0.06cm}
\begin{table}[!h]
\caption{\label{table_} Reconstruction of phases in cases with $N=1$.}
\begin{center}
\begin{tabular}{|c|c|c|c|c|c|c|c|c|}
\hline
Ex. & \multicolumn{2}{|c|}{Original phases} & $q(0,0)$ & \multicolumn{2}{|c|}{Coeff. of $F(z)$} & Aux. spectrum & \multicolumn{2}{|c|}{Recovered phases} \\
\hline
 & $\phi_0$ & $\phi_1$ & $q$ & $a_0$ & $a_1$ & $\mu$ & $\phi_0$ & $\phi_1$ \\
\hline
1 & $0.4$ & $0.8$ & $1.5844 -
1.0839i$ & $ -0.4873$ & $-1.5561$ & $0.7780 +
0.3163i$ & $0.4005$ & $0.7995$ \\
\hline
 2 & $3.5416$ & $3.9416$ & $-1.5844 +
1.0839i$ & $ 0.4873$ & $1.5561$ & $-0.7780 -
0.3163i$ & $3.5420$ & $3.9411$ \\
\hline
3 & $0.4$ & $3.9416$ & $0.1463 + 0.2139i$ & $1.3218$ & $-1.5561$ & $0.7780 - 3.9526i$ & $0.4000$ & $3.9416$ \\
\hline
\end{tabular}
\end{center}
\end{table}

\subsection{Case of $N=2$}

In order to provide an example,
where $R(z)$ has poles that have to be considered in the phase reconstruction algorithm, we choose a case with $N=2$.
 
Let $z_0=-1+3 i$, $z_1=5 i$, $z_2=1+3 i$. Ref.~\cite{smirnov2013periodic} shows
that the associated  $q(x,t)$ is periodic in $x$.

Let $\phi_0=0.1$, $\phi_1=0.2$ and $\phi_2=0.3$.
Then the calculated  auxiliary spectrum consists of two points, $\mu_1=-2.1061 + 0.4161 i$ and $\mu_2=2.1061 + 0.4161 i$,  and 
$F(z)=z^3-38.4617 z$.

\begin{figure}[!h]
    \centering
    \includegraphics[scale=.8]{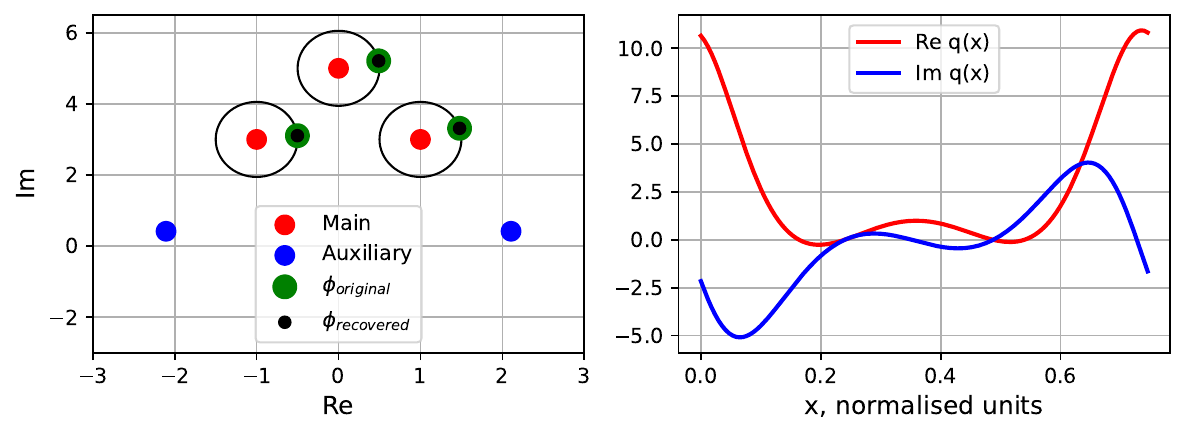}
    \caption{The example in the case with $N=2$ with main spectrum $z_0=-1+3 i$, $z_1=5 i$, $z_2=1+3 i$ and phases $\phi_0$, $\phi_1$, $\phi_2$.}
    \label{fig:2genus}
\end{figure}

In this case, both  $\mu_1$ and $\mu_2$ turn to be the poles of $R(z)$
and thus, we  have to proceed using \eqref{P-poles}
for calculating  $J_{00}(z)$. Then, the reconstruction gives 
$\phi_0=0.1007$, $\phi_1=0.2000$ and ${\phi_2=0.2993}$,
which is in good agreement with the original phases. The respective results are depicted in Fig.~\ref{fig:2genus}.

\enlargethispage{20pt}

\section{Conclusion}\label{sec:concl}

A finite-band (finite-genus) solution of the nonlinear Schr\"odinger
equation (in particular, its focusing version) can be characterized in terms of the solution
of a Riemann--Hilbert problem specified by (i) the set of endpoints of arcs constituting 
the contour for the RH problem and (ii) the set of real constants (phases), each being associated
with a particular arc. In the present paper we address the problem that can be 
described as ``an inverse problem to the inverse problem'', namely, given the finite-band
solution, generated via the solution of the RH problem and specified by a particular set of phases (assuming that the 
contour endpoints are fixed and that they are such that the 
finite-band solution is periodic in $x$) and evaluated as a 
function of $x$ for some fixed $t$, retrieve the phases.
Our approach is based on a sequence of consecutive transformations of the RH problem
characterizing the solution of the Cauchy problem for the NLS equation
in the periodic setting. Particularly, the role of the auxiliary spectrum points
in the RH formalism is clarified. 

\medskip

\noindent \textbf{Data accessibility.} The data and codes for the figures are available from the GitHub repository: https://github.com/Stepan0001/RHP-Direct-problem.git \cite{ourGit}.

\medskip

\noindent \textbf{Acknowledgments} 
S. Bogdanov and J. E. Prilepsky acknowledge the support from Leverhulme Trust, Grant No. RP-2018-063.


\vskip2pc

\bibliographystyle{RS}
\bibliography{shepelsky_etal}

\end{document}